\documentclass [amssymb, amsmath, pre,nofootinbib]{revtex4}

\usepackage{dcolumn}% Align table columns on decimal point
\usepackage[all]{xy}
\usepackage{subfigure}
\usepackage{color}

\usepackage{graphicx}
\usepackage{amsmath,amssymb,bm}

\begin{document}

\title{Non-Equilibrium Thermodynamics and Topology of Currents}

\author {Vladimir Y. \surname{Chernyak$^{a,b}$}}
\email{chernyak@chem.wayne.edu}
\author{Michael \surname{Chertkov$^b$}}
\email{chertkov@lanl.gov}
\author{Sergey V.  \surname{Malinin$^a$}}
\email{malinin@chem.wayne.edu}
\author{Razvan  \surname{Teodorescu$^b$}}
\email{razvan@lanl.gov}

\affiliation{$^a$Department of Chemistry, Wayne State University,
5101 Cass Avenue, Detroit, MI 48202}
\affiliation{$^b$Theoretical Division and Center for Nonlinear Studies,
LANL, Los Alamos, NM  87545}
\date{\today}

%\pacs{nn.mm.xx}{Third pacs description}

\begin{abstract}
In many experimental situations, a physical system undergoes stochastic evolution which may be
described via random maps between two compact spaces. In the current work, we study the applicability
of large deviations theory to time-averaged quantities which describe such stochastic maps,
in particular time-averaged currents and density functionals. We derive the large deviations principle for these
quantities, as well as for global topological currents, and formulate variational,
thermodynamic relations to establish large deviation properties of the topological currents.
We illustrate the theory with a nontrivial example of a Heisenberg spin-chain with a
topological driving of the Wess-Zumino type. The Cram\'{e}r functional of the topological
current is found explicitly  in the instanton gas regime for the spin-chain model in the weak-noise limit.
In the context of the Morse theory, we discuss a general reduction of continuous stochastic
models with weak noise to effective Markov chains describing transitions between stable fixed points.

\end{abstract}

%\pacs{02.50.Ey}{Stochastic processes}
%\pacs{02.50.Ga}{Markov processes}

\keywords{Non-equilibrium statistical mechanics, Topological Field Theory,
Cram\'er functional, topological current, Weak Noise, Instanton, Markov Chain}

\maketitle

\section{Introduction}
\label{sec:intro}

Dynamics of complex systems, in the presence of disorder and under
external forces, is often modeled by non-equilibrium stochastic processes.
Despite the considerable interest in such models, exact results,
or  even effective approximation methods are not readily available for
generic situations. Instead, specific results were derived under restricting
assumptions; for instance, the case of steady-states was investigated in
many publications (a complete list would be prohibitively long),
drawing upon the methods provided by the Large Deviations Principle
(LDP),  developed, for example, in Refs. \onlinecite{DV, BADZ1, BADZ2, BADZ3, DEHP}.

More recently, the LDP-based methods were applied to the case where the quantities of interest are
time-averaged observables, like currents or densities of particles
\cite{nu1,nu2,nu3,nu4,08MaesNW,08MaesN,07CCMT,07Derrida,GKP06}.
Even more attention was paid to the production of entropy,
which is actually a linear functional of currents, and related fluctuation theorems.
(See, for example,
Refs. \onlinecite{93EvansCohenMorris,95GallavottiCohen,97Jarz,00Crooks,05Seifert,
06CCJarz,05CCJarz,LS99,98Kurchan,HSch07,TurCCP07,07CKLTur}).
However, prior studies have only considered processes in spaces
with a trivial topological structure, in the sense that we will explain in the following.
To the best of our knowledge, the topological nature of global currents (fluxes)
has not been discussed before.

In most cases, stochasticity of dynamics and observation errors make detailed knowledge
of system trajectories unnecessary and distracting. Fortunately, one can often find
topological characteristics of the system, which are easier to observe.
We suggest that properly defined stationary currents are such topological characteristics.

In the present paper, extending on our recent preprint \cite{07CCMT},
we address the problem of deriving statistical properties of
%*
empirical
(time-averaged observable)
currents for non-equilibrium stochastic
processes which are equivalent to random maps between compact spaces
with nontrivial topology. We employ a method inspired by the LDP,
but also special field-theoretical tools developed originally in the
context of nonlinear sigma-models \cite{Abdalla}.

Our main results  concern deriving via LDP explicit asymptotic expressions for joint
distribution function of current density and density and distribution function
of total topological current in non-equilibrium steady state stochastic systems (stochastic maps)
defined on compact spaces. Existence of a nontrivial large deviations distribution for net
currents and the resulting thermodynamic relations are intimately related to
nontrivial configurations of maps from the base domain space to the target space.
We mainly focus on yet unexplored relationship between the large-deviations
probabilistic techniques for non-equilibrium systems and the topological structure
of the configuration spaces of the model. From that perspective, this article makes a
completely novel contribution. In the present study we consider a global static violation of detailed balance,
in contrast to stochastic pumping problems, where changing parameters in time
brings the system out of equilibrium \cite{07AstumianPNAS,08Sinitsyn,08RHJarz}.

%%%%%%%%%%%%%%%%%%%%%%%%%%%%%%%%%%%%%%%%%%%%%%%%%%%%%%%%%%%%%%%%%%%%
\begin{figure}[tp]
  \begin{center}
      \includegraphics[width=0.48\textwidth]{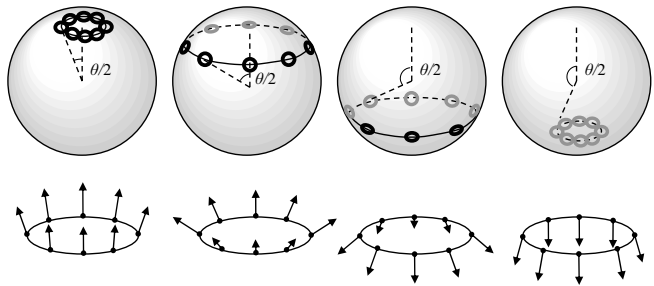}
  \end{center}
  \caption{Top: consecutive configurations of the string as it wraps around the sphere.
  Bottom: the same configurations in the discretized representation of the string by a cyclic spin chain.}
  \label{stringchain}
\end{figure}
%%%%%%%%%%%%%%%%%%%%%%%%%%%%%%%%%%%%%%%%%%%%%%%%%%%%%%%%%%%%%%%%%%%%%

The power of the aforementioned general results and technique is illustrated
on the enabling example of a circular spin chain, corresponding to a stochastic process mapping
from the torus to the sphere. (See Fig.~\ref{stringchain} for illustration.)
A physical realization of this model could be the following: consider a nano-structure represented
by a circular spin-chain of $N \gg 1$ classical interacting spins characterized by unit vectors
${\bm n}_i$ where $i= 1,\ldots,N$,
coupled to a stochastic external magnetic field $\bm B$. Note that such a device
could be used as a magnetic field detector, by measuring the response of the spin chain as a
function of time. It is most useful to consider the low-energy, long-wavelength limit of the
problem, which would correspond to a high sensitivity of the device.
Therefore, we will restrict our attention only to the spin couplings relevant to the long-wavelength approximation.
The spin system can be driven in other ways than by an external magnetic field.
For example, a nonconservative force, a so-called spin transfer torque,
is caused by a spin-polarized current \cite{spintorque1,spintorque3}.

Other possible realizations of the spin-chain model are
molecular motors \cite{Astumian,F1ATPase,RMPmotors}
which often exhibit periodic motions resulting from nonconservative
driving. Although specific examples may not have been yet discovered,
we believe that there are relatively simple non-equilibrium bio-molecular systems whose
functioning is controlled by topological currents,
probably far more complex than those discussed here in the
context of the spin-chain model.

Performing precise measurements in such setups, in the presence of fluctuations, is essentially related to the
ability to detect collective modes of the chain ${\bm n}_i$, by integrating
the response over a time interval. Therefore, such a device would be
a natural detector of the total {\emph{current}} associated with the
integrated response of the entire chain.
%*
Our analysis will focus on the probabilistic description of the
%which will describe distribution of the
empirical \emph{current} (time-integrated response)
and generalize to compact spaces the study of the
current {\emph{density}} distribution (or a 1D current), the focus of earlier works
\cite{nu1, nu2, nu3, nu4, 08MaesNW, 08MaesN, 07CCMT, 07Derrida, GKP06} in both single-
and many-particle systems.
The total current in our spin-chain example originates from global, topological characteristics of the system,
as locally the stationary force driving the system is potential.
In fact, this example represents the case of the  topological driving associated with
a multi-valued potential of the Wess-Zumino type
\cite{82Wit,82Nov,Abdalla},  and thus the essential part of our analysis will be
devoted to establishing large deviations characteristics of
the global (topological) current in the spin-chain model.

The material in the paper is organized as follows. In Section \ref{sec:Mod+Res}
we introduce a general stochastic model and a specific example with a nontrivial topology.
All the results of the paper are briefly discussed and listed at the end of this introductory Section.
In Section \ref{sec:currents-cycles} we present a topological picture of stochastic currents.
In Section \ref{sec:LDP} we consider overdamped continuous stochastic processes.
We review some results related to statistics of the empirical density,
and current density, as well as other extensive observables derived from them.
We also extend the large deviation theory to nontrivial topologies and
develop a general approach to obtain large deviation functions for topological currents.
Section \ref{sec:spin-chain} is devoted to a non-equilibrium stochastic model with a nontrivial
topology: a classical Heisenberg spin chain driven by a topological term of the Wess-Zumino type.
In Section \ref{sec:conclusions}, in addition to giving a summary and conclusions,
we discuss a general reduction of continuous stochastic models in the weak-noise limit to
effective Markov chains with the help of the Morse theory.
Appendices contain some technical details and auxiliary material.

\section{Models and Statements of the Results}
\label{sec:Mod+Res}

\subsection{General Stochastic Model}
\label{subsec:Gen}

We consider a stochastic process/trajectory ${\bm\eta}_{t}=\{
{\bm\eta}(\tau) | 0 \le \tau \le t \}$, or more rigorously
${\bm\eta}_{t}:[0,t] \to M$ of duration $t$, which occurs in a configuration
space M, i.e., ${\bm\eta}(\tau) \in M$ for $0 \le \tau \le t$.
The configuration space $M$ is assumed to be a manifold of dimension
$m={\rm dim}M$, so that the particle position ${\bm\eta}(\tau)$ at any given
time can be characterized by a set $\eta^{i}(\tau)$ of local coordinates
with $i=1,\ldots,m$. Our stochastic process can be described by the
following Langevin equations
\begin{eqnarray}
\label{Ito-eq}
&& \dot\eta^{i}(\tau)=F^{i}({\bm\eta})+\xi^{i}(\bm\eta, \tau)\,,
\end{eqnarray}
%*
which is a continuous limit of the well defined discrete-time stochastic differential
equations written, for example, in the It\^o form.  The quantities
discussed below that depend on the continuous time
should be understood as limits of the their properly discretized forms.
In Eq. (\ref{Ito-eq}) $F^{i}$ denotes
the deterministic (advection) component of the particle velocity, linearly related to the
driving force $F_{j}$ (overdamped dynamics),
\begin{eqnarray}
\label{force-velocity}
&&
F^{i}=g^{ij}F_{j}\,,
\end{eqnarray}
via the mobility tensor $g^{ij}(\bm\eta)$ that can be viewed as a
Riemann metric on the configuration space $M$. Due to the Einstein relation
(fluctuation-dissipation theorem), the same tensor $\kappa g^{ij}$, weighted with a factor $\kappa$
that controls the noise strength, characterizes the
correlations of the Gaussian Markovian noise in Eq.~(\ref{noise}):
\begin{eqnarray}
\label{noise}
&& \langle \xi^{i}(\bm\eta,\tau_{2}) \xi^{j}(\bm\eta,\tau_{1}) \rangle=\kappa g^{ij}(\bm\eta) \delta
(\tau_{2}-\tau_{1})\,.
\end{eqnarray}
If our stochastic
dynamics is interpreted as a result of elimination of fast components in harmonic bath modeling (to
achieve the Markov limit), the Einstein relation means that the bath is at equilibrium at
temperature $\kappa$, and non-equilibrium features of the system's stationary state can result only
from the non-potential nature of the driving force ${\bm F}$. Hereafter we imply summation over
the repeating indices and assume, without loss of generality, that the metric is curvature-free.
Eqs.~(\ref{noise}) are consistent with
the stochastic (Onsager-Machlup \cite{OM53}) action
%-------------------------------------------------------------------------------------
\begin{eqnarray}
\label{action} S( {\bm\eta}_t) = \frac{1}{2\kappa}\int_{0}^{t}d{\tau} \,
g_{ik}\left(\dot{\eta}^{i}-F^{i}({\bm \eta})\right)
\left(\dot{\eta}^{k}-F^{k}({\bm \eta})\right)
\end{eqnarray}
%-------------------------------------------------------------------------------------
(with $g_{ij}$: $g_{ij}g^{jk}=\delta_{i}^{k}$)
defining the probability measure over ${\bm\eta}_t$, such that the stochastic average of a functional
$\bullet({\bm\eta}_t)$ of ${\bm \eta}_t$
(e.g., an observable accumulated over time $t$) is evaluated according to
\begin{eqnarray}
\langle \bullet({\bm \eta}(t)\rangle_\xi
=\frac{\int_M{\cal D}\bm\eta(\tau) \bullet(\bm\eta_t) \exp(-S( {\bm\eta}_t))}{
\int_M{\cal D}\bm\eta(\tau) \exp(-S( {\bm\eta}_t))},
%\quad {\cal D}\eta(\tau)=\prod_{\tau}d\eta(\tau),
\label{av_eta}
\end{eqnarray}
where the denominator is usually called the partition function. In Eq.~(\ref{av_eta})
we use standard notations for the path integrals over trajectories
and assume proper discretization over the time interval $[0,t]$.
%*
Distinction between different discretization conventions is irrelevant for the Cram\'er functions
in the low-noise limit.

We introduce the
%time-averaged
empirical density and current at a point ${\bm x}$ of the trajectory's
configuration space:
\begin{eqnarray}
 & \rho_{t}({\bm\eta}_t,{\bm x})\equiv t^{-1}\int_{0}^{t}d\tau \delta({\bm x}-{\bm\eta}(\tau)),
 \label{rho-jay1} \\
 & J_{t}^{i}({\bm\eta}_t,{\bm x})\equiv t^{-1}\int_0^t d \tau \dot{\eta}^{i}
\delta({\bm x}-{\bm\eta}(\tau)).
\label{rho-jay2}
\end{eqnarray}
We are interested in the large-deviation limit of the joint probability distribution function for
$\rho_{t}$ and ${\bm J}_{t}$,
\begin{align}
\label{prob} \mathcal{P}_{t}({\bm J},\rho)\equiv \langle \delta(\rho_{t}-\rho)
\delta({\bm J}_{t} - \bm{J})\rangle_{\xi} \,.
\end{align}

Assuming that the observation time $t$ is large and focusing primarily on statistics of $\rho_t,
{\bm J}_t$ defined above one observes that a distinction between an open
trajectory with ${\bm\eta}(0)\neq{\bm\eta}(t)$ and a closed trajectory with
${\bm\eta}(0)={\bm\eta}(t)$ disappears at $t\to\infty$. This fact, also discussed in detail in
Section \ref{sec:currents-cycles}, allows us to focus on the analysis of closed
trajectories.

The joint distribution function of current density and density defined in Eq.~(\ref{prob}) is a
very useful and rich object carrying sufficient amount of dynamical and topological information
about the system one would normally be interested in.  However, from the point of view of
experimentally and computationally desirable low-dimensional characterization of the stochastic
system, the functional $\mathcal{P}_{t}({\bm J},\rho)$ is still too complicated.
One would like to introduce a version of
Eq.~(\ref{prob}) of smaller dimensionality, with inessential parameters integrated out.  We suggest that a
{\it topologically protected} quantity satisfying these requirements is the equivalence class components
of the intersection index between a closed trajectory ${\bm\eta}_t$ and a cross-section $\alpha$ of
$M$, which we denote symbolically as
\begin{eqnarray}
\omega_t^{[\alpha]}=\int_\alpha {\bm J}_t.
\label{index}
\end{eqnarray}
Here $[\alpha]$ indicates that the object is invariant with respect to continuous transformation
from one cross-section to another within the same equivalence class. Obviously, currents can be
added and multiplied by numbers, where respective operation is executed over the corresponding current densities.
Therefore, currents ${\bm\omega}_{t}$ can be viewed as vectors that reside in a certain vector space,
a current space. The current components $\omega_{t}^{[\alpha]}$, introduced
in Eq.~(\ref{index}), are labeled by linearly independent equivalence classes $[\alpha]$ of cross
sections, whose number defines the dimension of the current space. Further details (including
formal definitions) will be given in Section \ref{sec:currents-cycles}.
As in other parts of this manuscript, our main focus will be on evaluating the LDP asymptotic
for the respective distribution function
\begin{eqnarray}
\label{probI} \mathcal{P}_{t}(\bm{\omega})\equiv \Biggl\langle
\prod_{[\alpha]}\delta(\omega_t^{[\alpha]}-\omega^{[\alpha]})\Biggr\rangle_{\xi}.
\end{eqnarray}
Notice that the number of the linearly independent equivalence classes $[\alpha]$
(the dimension of the current space) naturally depends on $M$, being a small number for
common topological problems (field theories). For example, for the
problem considered in Section \ref{sec:spin-chain} the current space
is one-dimensional.

\subsection{Circular Spin Chain}
\label{subsec:spin-chain}

In the large $N$ limit, $N \to \infty$, we can describe the circular spin-chain model using a map
$\bm{n}(y,\tau)=(n^{a}(y,\tau)|{a=1,2,3}\ \&\ \sum_{a} n^{a} n^{a}=1)$, which represents the
three-dimensional unit vector parameterized by the angle $y\in [0, 2\pi]$ at the time $\tau$. By
imposing periodic boundary conditions in $y$ and $\tau$,
we also find convenient to think about the model as of a $(1+1)$ field
theory, i.e. as of a stochastic map $S^{1}\times S^{1} \to S^{2}$.

To illustrate the
general topological results we will consider the following
version of the model (\ref{Ito-eq}):
%-------------------------------------------------------------------------------------
\begin{eqnarray}
\label{eq-determ-explicit}
&& \partial_{\tau}{\bm n}={\bm F}({\bm n})+\bm\xi, \;\;\;
{\bm F}({\bm n})={\bm F}_{\rm e}({\bm n})+u[{\bm n},\partial_{y}{\bm n}], \;\;\;
{\bm F}_{\rm e}({\bm n})=
v\left(\partial_{y}^{2}{\bm n}+(\partial_{y}{\bm n}\cdot\partial_{y}{\bm n}){\bm n}\right), \\
&&  \langle
\xi^{a}(y_{1};\tau_{1})\xi^{b}(y_{2};\tau_{2})\rangle=\kappa\delta(\tau_{1}-\tau_{2})\delta(y_{1}-y_{2})
\left(\delta^{ab}-n^{a}(y_{1},\tau_{1})n^{b}(y_{1},\tau_{1})\right),      \label{n-noise}
\end{eqnarray}
where parameters $u$ and $v$ correspond to strength of driving and overall noise normalization
respectively and $[\bullet,\bullet]$ is the standard notation for the vector cross-product.
(Eq.~(\ref{n-noise}) guarantees the transversality of the noise term to the ${\bm n}(y,t)$ field,
and it is also straightforward to verify that all terms on the rhs of Eq.~(\ref{n-noise}) are in
fact transversal to ${\bm n}(y,t)$. As argued in the Introduction, this model can describe the
long-wavelength limit of a nano-scale spin device manipulated by magnetic field \cite{spintorque1,spintorque3}.
In fact, Eqs.~(\ref{eq-determ-explicit})
represent the most general weak-noise, weak-driving stochastic equations which may be built for the
map $\bm{n}(y,t)$ in the long-wavelength limit, thus keeping only low-order spatial gradients (over
$y$).
It is important to emphasize that the local force in the spin-chain model is conservative,
and the non-equilibrium character of the stochastic process is due to global
(topological) effects. These topological aspects of the model will be discussed
in detail in the two first Subsections of Section \ref{sec:spin-chain}.

The Onsager-Machlup action (\ref{action}) that corresponds to the Langevin dynamics described by
Eqs.~(\ref{eq-determ-explicit}) and (\ref{n-noise}) has a form
\begin{eqnarray}
\label{S-OM-sigma-full} S({\bm
n})=\frac{1}{2\kappa}\int_{0}^{t}d\tau\int_{S^{1}}dy\left(\partial_{\tau}{\bm n}-u[{\bm
n},\partial_{y}{\bm n}]-v\left(
\partial_{y}^{2}{\bm n}+(\partial_{y}{\bm
n}\cdot\partial_{y}{\bm n}){\bm n}\right)\right)^{2}
\end{eqnarray}
Setting $v=0$ turns this action into the known model, often called a $(1+1)$ nonlinear $\sigma$-model on a sphere with a topological term (often referred to as a $\theta$-term) \cite{Abdalla}:
\begin{eqnarray}
\label{S-OM-sigma-u-only} \left.S({\bm
n})\right|_{v=0}=\frac{1}{2\kappa}\int_{0}^{t}d\tau\int_{S^{1}}dy\left((\partial_{\tau}{\bm
n})^{2}+u^{2}(\partial_{y}{\bm n})^{2}-2u({\bm n}\cdot[\partial_{\tau}{\bm n},\partial_{y}{\bm
n}])\right).
\end{eqnarray}
The nonlinear $\sigma$-model of Eq.~(\ref{S-OM-sigma-u-only}), considered as an imaginary-time field theory, has
ultraviolet divergences, and thus requires a small scale regularization. In fact, our circular spin
chain model (\ref{eq-determ-explicit}) may be viewed as a regularized counterpart of the
$\sigma$-model, where the number $N$ of spins plays the role of the ultraviolet cut-off parameter.
An intuitive explanation for this regularization is that the $v$-term in
Eq.~(\ref{eq-determ-explicit}) acts to align all the spins and, therefore,
suppresses the short-range fluctuations.

In this manuscript we will focus on analysis of the circular spin-chain model in the weak-noise
limit $\kappa \ll |u| \lesssim v$, which will also be coined (by the reason to be spelled out later in
Section \ref{sec:spin-chain}) the limit of ``instanton gas", correspondent to moderate topological
driving \footnote{Another ``anomalous" regime  of the sigma-model, $v \ll \kappa \ll |u|$, was
analyzed by Polyakov and Wiegman \cite{PolykovWiegmann1,PolykovWiegmann2} in the zero topological charge
sector (not relevant for our application). The authors mapped the model onto a system of infinite-component
massless interacting fermions, analyzed it with the Bethe ansatz approach and thus arrived at a
remarkable exact and nontrivial solution. Notice, that an intermediate case of ``interacting
instantons", correspondent to strong topological driving and vanishing noise, $\kappa \ll v \ll
|u|$, constitutes yet another interesting regime, which to the best of our knowledge was not
studied yet. The Cram\'er function in this later case can be calculated explicitly by integrating
over the instanton solutions and Gaussian fluctuations around them in the limit $v=0$. As
demonstrated in Ref. \cite{79FFS} in the context of
the $\sigma$-model as a field theory, the problem is equivalent to finding the ground-state energy
of the corresponding $(1+1)$ sine-Gordon model, where $v$ will play a role of the ultraviolet
cut-off parameter.
%*
In the case $v<u$ the saddle and stable points of the spin-chain system coalesce producing
a peculiar fixed point at $\theta=0$ which is stable when approached from one direction and unstable
from the other one. All such situations are beyond the scope of this work.}.
We demonstrate that the configuration space of the spin-chain model has one
topologically nontrivial cycle (i.e., the current $\omega$ is single-component). In the considered
limit the system spends most of the time around its stable configuration, whereas the current is
generated by rare events, referred to as instantons, whose interaction can be neglected due to
long time intervals between them.

\subsection{Statement of Results}
\label{subsec:StatRes}

The main results reported in this manuscript are as follows:
\begin{itemize}

\item
In Section \ref{sec:currents-cycles} we establish a topological nature of the average current
generated over a long time in a stochastic system. The topological stochastic current ${\bm\omega}$
resides in a vector space, referred to as the current space,
whose dimension is given by the number of independent $1$-dimensional cycles of the system
configuration space $M$.
We demonstrate that there are two equivalent ways to view the generated topological currents:
(i) the rates with which the stochastic trajectory
loops around the independent $1$-cycles, and
(ii) the equivalence
classes of the divergence-free current density distributions. The equivalence of these two views is
established by the Poincar\'e duality represented by Eq.~(\ref{Poincare-duality}).

\item
We show in Section \ref{sec:LDP} that
for a stationary stochastic process
the joint probability distribution for the
%time-averaged
empirical current density and density
in the $t\to\infty$ limit takes the form
$\mathcal{P}_{t}(\bm{J}, \rho) \sim\exp (-t\mathcal{S}(\bm{J},\rho))$
with the Cram\'er functional
%-------------------------------------------------------------------------------------
\begin{align}
\label{CDDF-continuous-explcit} {\cal S}({\bm J}, \rho )=\int_{M}d{\bm x} \frac{
(\bm{F}\rho-\bm{J}-(\kappa/2)\bm{\partial}\rho)^2}{2\kappa\rho} .
\end{align}
This generalizes the previously reported results \cite{
nu1, nu2, nu3, nu4, 08MaesNW, 08MaesN, 07CCMT} to the case of topologically nontrivial compact spaces.

\item
In Section \ref{subsec:general-topo-currents-functional-derivation} we describe a general method
to derive large-deviation statistics of particular currents (e.g., the Cram\'{e}r function
of the topological currents) from Eq. (\ref{CDDF-continuous-explcit})
in a variational way by solving a respective set of equations for currents and densities.

\item
We illustrate the utility of the general approach on the example of the thermodynamic limit $N\to
\infty$ of the spin-chain model defined above in Section \ref{subsec:spin-chain}.
We show that the topological current space of the model is one-dimensional, i.e., the generated current
is described by $\omega\in \mathbb{R}$. We compute the Cram\'er function ${\cal S}(\omega)$
in the weak-noise limit $\kappa
\ll v,|u|$ for not too large values of the generated current $\omega$ for $|\omega|\ll
|v|(\ln(|v|/\kappa))^{-1}$, when the instanton gas mechanism dominates the current generation.
In this instanton-gas regime the Cram\'er function ${\cal S}(\omega)$
has the same form as for the
effective $1D$ random walk (equivalent to a circular two-channel single-state Markov chain)
with jump rates $\kappa_+$ and $\kappa_-$ in opposite directions:
%-------------------------------------------------------------------------------------
\begin{align}
\label{S-omega-explicit} & {\cal S}(\omega)=\kappa_{+}+\kappa_{-}+\omega\ln\frac{\omega
+\sqrt{\omega^{2}+4\kappa_{+}\kappa_{-}}}{2\kappa_{+}}-\sqrt{\omega^{2}+4\kappa_{+}\kappa_{-}} \,,
\end{align}
%-------------------------------------------------------------------------------------
the rates $\kappa_{\pm}$ being expressed through the parameters $u$ and $v$ of the model
(\ref{eq-determ-explicit}) by means of Eqs.~(\ref{k0}), (\ref{kappapm-S12}), (\ref{detdet}),
(\ref{int-zero-modes-G}), and (\ref{S1S2}).

\item
In general, the vector of topological currents ${\bm \omega}$ is not related
to the work produced by the driving force, $\int_0^td\tau ({\bm F}({\bm\eta})\cdot \dot{\bm\eta})$.
However, in the case of topological driving, as in the spin-chain model,
the work becomes a linear functional of the current vector.

\end{itemize}

\section{Topological view of stochastic currents}
\label{sec:currents-cycles}

\begin{figure}[ht]
  \begin{center}
      \includegraphics[width=0.35\textwidth]{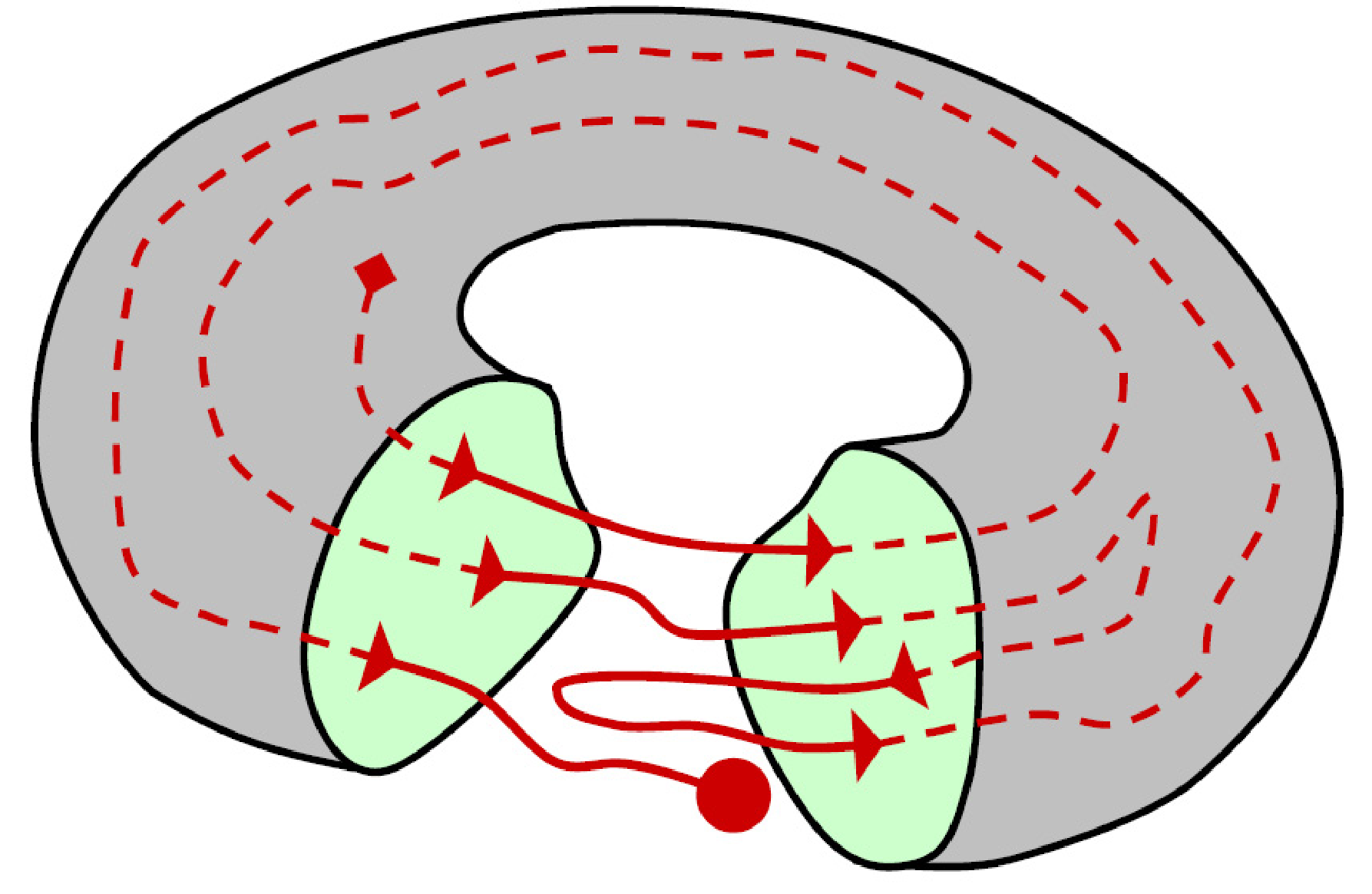}
  \end{center}
  \caption{A single open trajectory in the circular wire.
  }
  \label{fig:wire}
\end{figure}

Historically, a concept of current appeared in physics on a macroscopic level as a way to describe
a flux of any kind.
%of a kind.
Consider an electric circuit, represented by a circular wire with a static electric field
(for example, provided by a battery) where the electrons on average move in one direction.
The current is defined as a charge crossing some oriented section $\alpha$ per unit time. The
current can be considered as a sum of currents from individual particles. An individual
contribution $\omega$ is given by $\omega=Nt^{-1}$ with $N=N_{+}-N_{-}$, where $N_{\pm}$ is the
number of times the particle trajectory ${\bm\eta}$ crosses the section $\alpha$ in the positive
and negative directions, respectively (see Fig. \ref{fig:wire}).
%----------------------------------------------------------------------
%----------------------------------------------------------------------

The configuration space $M$ for the particles in the circular wire is three-dimensional, i.e.,
$m={\rm dim M}=3$, and can be represented as $M=S^{1}\times D^{2}$ (with $S^{1}$ and $D^{2}$ being
a circle and a two-dimensional disc, respectively), whose boundary is $\partial M=S^{1}\times
S^{1}$. If the particle trajectory ${\bm\eta}$ is closed, i.e., ${\bm\eta}:S^{1}\to M$, the number
$N$ does not change upon deformations of the trajectory and the cross-section $\alpha$.
This number, known as the {\it intersection index},
only depends on the equivalence classes
$[{\bm\eta}]$ and $[\alpha]$ and can be denoted by $[{\bm\eta}]*[\alpha]\in \mathbb{Z}$. The
equivalence, based on deformations, is the {\it homotopical} equivalence \footnote{Here and below
$[\cdot]$ is used as a notation for the equivalence class of $\cdot$.}.

\subsection{Intersection Index and Stochastic Currents}
\label{subsec:current-inersect-index}

The intersection index does not change if any of the cycles is replaced by a {\it homologically}
equivalent counterpart. Since cycles can be added and multiplied by the integers (by forming
disjoint unions and changing orientations), this equivalence can be understood if we define a zero
cycle. Formally, a $j$-dimensional cycle is called homologically equivalent to zero, if it is a
boundary of a $(j+1)$-dimensional region mapped into $M$. Obviously homotopy equivalent cycles are
homologically equivalent. The set (actually an Abelian group) of homological classes of
$j$-dimensional cycles in $M$ is called the $j$-th homology of $M$ and denoted by
$H_{1}(M;\mathbb{Z})$. In the circular wire, an element of $H_{1}(M;\mathbb{Z})$ represents the
number of times the particle trajectory moves around the circuit: $H_{1}(M;\mathbb{Z})\cong \mathbb{Z}$.
%$[c]\in H_{1}(M;\mathbb{Z})$
The current ${\bm\omega}$ associated with the trajectory ${\bm\eta}$ can be naturally defined as
${\bm\omega}=[{\bm\eta}]t^{-1}\in H_{1}(M;\mathbb{R})=H_{1}(M;\mathbb{Z})\otimes_{\mathbb{Z}}
\mathbb{R}$. In the circular wire of Fig.~\ref{fig:wire} the current has only one component
because $H_{1}(M;\mathbb{R})=\mathbb{R}$.

The topological picture, presented above on the simple example of a circular
circuit, can be extended to a much less intuitive general case in a pretty straightforward way by
viewing an averaged current ${\bm\omega}$ generated in stochastic dynamics in the configuration
space $M$ as an element in $H_{1}(M;\mathbb{R})$, the first homology group of $M$ with real
coefficients, associated with the homology class $[{\bm\eta}]$ of a stochastic trajectory. Since
$H_{1}(M;\mathbb{R})$ is a real and in most relevant cases finite-dimensional vector space, the
generated current can be viewed as a vector. The homology groups can be finite-dimensional and
computable even in the field theory when the configuration space $M$ is represented by an
infinite-dimensional (in the non-regularized continuous limit) space of maps. In particular, in our
enabling  case of $M={\rm Map}(S^{1},S^{2})$ discussed in Section~\ref{sec:spin-chain},
the current also has only one component since $H_{1}({\rm Map}(S^{1},S^{2});\mathbb{R})\cong
\mathbb{R}$.

Notice also that, originating from counting, stochastic currents can be viewed as {\it
topologically protected observables}, available in single-molecule measurements, which provide
stable and at the same time rich information on the underlying stochastic processes.

The topological picture of stochastic currents can be formulated in the simplest way when defined
for closed trajectories (loops). However, a general stochastic trajectory is open, i.e.,
its end point is typically different from the starting one and thus adopting the topological
language of intersection indexes and currents may seem problematic.  In the long-time
limit considered in this manuscript, extension from open to close trajectories does not constitute
a problem. Indeed, when counting the intersection index $N$ of a long trajectory, one can always
close it with a segment (e.g., a geodesic line) that is much shorter than the trajectory itself,
which creates an uncertainty no more than one in a big number $N\gg 1$.
%Then,
%*
In contrast to a boundary effects with unbounded fluctuations \cite{06PRV,03vZC},
in the long-time limit
in our case
one can safely ignore changes in the statistical properties of the trajectories caused by
this modification.

\subsection{Currents, Current Densities, Zero-Curvature Vector Potentials and Poincar\'e duality}
\label{subsec:current-density-Poicare}

Currents can also be represented by the current densities ${\bm J}$ defined as time
integrals over trajectories as in Eq.~(\ref{rho-jay2}). In this Subsection
we will discuss some important ``static" relations between ${\bm J}$,  the current
${\bm\omega}$, and the intersection invariant $[\alpha]$.

The time-averaged current density from Eq.~(\ref{rho-jay2}) is
a random variable on the space of stochastic trajectories $\bm \eta(\tau)$ and satisfies the relation
\begin{eqnarray}
\label{div-J-t}
{\rm div}_{\bm x}{\bm J}_{t}({\bm\eta}_{t},{\bm x})=
t^{-1}\left(\delta({\bm x}-\eta_{t}(t))-\delta({\bm x}-\eta_{t}(0))\right).
\end{eqnarray}
Therefore, our first observation is that the random variable ${\rm div}{\bm J}_{t}\sim t^{-1}$
vanishes in the limit $t\to\infty$.
Moreover,  Eq.~(\ref{div-J-t}) guarantees that the current density is exactly divergence-free,
${\rm div}_{\bm x}{\bm J}_{t}({\bm\eta}_{t},{\bm x})=0$,
if only closed stochastic trajectories are considered,
which, as argued in Section \ref{subsec:current-inersect-index},
does not affect the long-time behavior of the relevant distributions.

The vector field of the current density on the $m$-dimensional phase space can be naturally viewed as
a differential form of rank $(m-1)$ because its integration over a $(m-1)$-dimensional
cross-section section $\alpha$ results in a current. In the following we will use the same
notations for vector fields and the corresponding differential forms. The divergence of the current
density ${\bm J}$ is represented by the exterior derivative as ${\rm div}_{\bm x}{\bm J}=d{\bm J}$,
and Eq. (\ref{div-J-t}) suggests that the current density is a closed from: $d{\bm J}=0$.
Therefore, it represents a class $[{\bm J}]$ in the de Rham cohomology $H^{m-1}(M;\mathbb{R})$. The
relation between the homological and current-density representations of the current is determined by
the Poincar\'e duality $H_{1}(M;\mathbb{R})\cong H^{m-1}(M;\mathbb{R})$ \cite{Spanier}. This can be
formally expressed as
\begin{eqnarray}
\label{Poincare-duality}
\forall [\alpha]\in H_{m-1}(M;\mathbb{Z}):\quad  {\bm\omega}*
[\alpha]=t^{-1}[{\bm\eta}]*[\alpha]=\int_{\alpha}{\bm J},
\end{eqnarray}
thus representing that per unit time intersection index
$t^{-1}[{\bm\eta}]*[\alpha]$ of a closed trajectory ${\bm\eta}$ with a cross-section $\alpha$ is
equal to the integral over $\alpha$ of the current density ${\bm J}$ produced by the trajectory.

We conclude this Section by noting that the Poincar\'e duality $H_{m-1}(M;\mathbb{R})\cong
H^{1}(M;\mathbb{R})$ in the complementary dimension leads to a natural representation of the
sections $\alpha$ in terms of the vector potentials ${\bm A}$, which are curvature-free, i.e.
\footnote{Where here and below $\partial_i$ is our shortcut notation for $\partial_{x_i}$.}
\begin{eqnarray}
(\partial_{i}A_{j}-\partial_{j}A_{i})dx^{i}\wedge dx^{j}=0,
\label{zero-curv}
\end{eqnarray}
since the cross-section $\alpha$ represents a homology class $[\alpha]\in H_{m-1}(M;\mathbb{R})$,
whereas the corresponding vector potential ${\bm A}$ represents a gauge equivalence class $[{\bm
A}]\in H^{1}(M;\mathbb{R})$. Here in Eq.~(\ref{zero-curv}) we used standard wedge-product notations
for  the differential $1$-forms \cite{G-H-alg-geom}. The relation ``$[{\bm A}]$ correspond to
$[\alpha]$ via the Poincar\'e duality'' can be conveniently expressed in a way extending
Eq.~(\ref{Poincare-duality}),
\begin{eqnarray}
\label{Poincare-duality-compliment}
\forall[{\bm J}]\in H^{m-1}(M;\mathbb{Z})| \, {\rm div}{\bm J}=0:\quad \int_{\alpha}{\bm J}=\int_{M}{\bm
A}\wedge{\bm J}=\int_{M}d{\bm x}{\bm A}\cdot{\bm J},
\end{eqnarray}
which implies that the current components ${\bm\omega}*[\alpha]$ can be labeled by the gauge
classes of the curvature-free vector potentials. This expresses the general Poincar\'e duality
$H_{j}(M;\mathbb{R})\cong H^{m-j}(M;\mathbb{R})$ in dimension $j=m-1$, i.e.,
$H_{m-1}(M;\mathbb{R})\cong H^1(M;\mathbb{R})$. Using Eq.~(\ref{Poincare-duality-compliment})
it can be formulated as follows. (i) With any $(m-1)$-cycle $\alpha$ we can associate a closed
$1$-form ${\bm A}$ so that Eq.~(\ref{Poincare-duality-compliment}) holds for any divergence-free
current density distribution ${\bm J}$. (ii) Any two forms ${\bm A}$ and ${\bm A}'$ that satisfy
the condition (i) are equivalent, $[{\bm A}]=[{\bm A}']$. (iii) Homologically equivalent cycles
generate equivalent forms, i.e., if ${\bm A}$ and ${\bm A}'$ are respectively generated by $\alpha$ and
$\alpha'$, then $[\alpha]=[\alpha']$ implies $[{\bm A}]=[{\bm A}']$. Note that the
conditions (i)-(iii) define a linear map $H_{m-1}(M;\mathbb{R})\to H^1(M;\mathbb{R})$. The Poincar\'e
duality also means that this map is an isomorphism. Note that
Eq.~(\ref{Poincare-duality-compliment}) is noticeably distinct from its
counterpart Eq.~(\ref{Poincare-duality}). The latter describes the Poincar\'e duality in the
complementary dimension $j=1$, by associating the currents ${\bm\omega}\in H_{1}(M;\mathbb{R})$
with the equivalence classes $[{\bm J}]\in H^{m-1}(M;\mathbb{R})$ of divergence-free current
densities ${\bm J}$.

%\section{LDP for time-averaged stochastic currents}
\section{LDP for empirical currents}
\label{sec:LDP}

In this Section, we describe results concerning the Large Deviations Principle (LDP) for (possibly)
dependent sequences of random variables, including its application to the case of time-averaged
current densities, and present a path-integral derivation for the relevant functionals. The
material of this Section is organized as follows. In subsection~\ref{subsec:LDP-review} we very
briefly review the foundations that stand behind the LDP and formulate a variational principle
relating the LDP for current density and density to the LDP for conjugated vector and scalar potentials. In
Subsection~\ref{subsec:LDP-path-integral} we describe a convenient representation of the LDP in the
intuitive path-integral language.
In Subsection~\ref{subsec:general-topo-currents-functional-derivation} we present a general method
to derive the Cram\'{e}r function of the topological currents from the Cram\'{e}r functional of
the joint density and current density distribution.

%\subsection{LDP for time-averaged stochastic currents and the G\"artner-Ellis theorem}
\subsection{LDP for empirical currents and the G\"artner-Ellis theorem}
\label{subsec:LDP-review}

G\"artner-Ellis (G-E) theorem \cite{G-E1,G-E2}  formalizes the Large Deviation Principe (LDP). It
provides a convenient theoretical tool for studying the long-time behavior of stationary driven
systems. In this Subsection we discuss fundamental relations and objects associated with the G-E
theorem and the LDP.

Consider a sequence of random variables  ${\bm\xi}_\tau\in {\cal H}$ with $\tau=1,2,\ldots,t$.
Denote by $P_t({\bm \varphi})$ the distribution of the average
${\bm\varphi}_t = t^{-1}\sum_{\tau=1}^t {\bm\xi}_\tau$ on the vector space ${\cal H}$.
Define the corresponding generating function
${\cal Q}_{t}({\bm\psi})=\left\langle \exp\left( t{\bm\psi}\cdot{\bm\varphi}_{t}\right)\right\rangle$,
with ${\bm\psi}\in {\cal H}^{*}$ being linear functionals in ${\cal H}$.
%Let us also assume that the subscript $t$ marks the final time slice of the
%discretized time $t=1,2,\ldots$.
If the limit
\begin{eqnarray}
\label{fenergy}
{\lambda}({\bm\psi}) = \lim_{t \to \infty} t^{-1}\ln\left({\cal Q}_{t}({\bm\psi})\right)
\end{eqnarray}
does exist,
is represented by a convex and bounded from below function, and
\begin{equation}
\label{ratefunction}
{\cal S}({\bm\varphi})=\sup_{{\bm\psi}}\left({\bm\psi}\cdot{\bm\varphi}-\lambda({\bm\psi})\right)
\end{equation}
is represented by a bounded from above lower semi-continuous
function with compact level sets, then $\{{\bm\varphi}_{t}\}$ satisfies the Large Deviations
Principle (LDP) with the \emph{rate} (or \emph{Cram\'er}) \emph{function} ${\cal S}$.
This is known as the G\"artner-Ellis (G-E) theorem \cite{G-E1,G-E2}.
The LDP is formulated in terms of the probabilities $P_{t}(K)=\int_{K}dP_{t}({\bm\varphi})$ for
the random variable $\{{\bm\varphi}_{t}\}$ to belong to $K\subset {\cal H}$.
In many cases it can be formulated in a
stronger, yet simpler form (see \cite{G-E1,G-E2} for a general formulation): for any (measurable) set
$K\subset {\cal H}$
\begin{eqnarray}
\label{ge} \lim_{t\to \infty}t^{-1}\ln\left(P_{t}(K)\right)=-{\cal S}(K),
\end{eqnarray}
where by definition the Cram\'er function is
\begin{eqnarray}
\label{define-Q-s}{\cal S}(K)\equiv \inf_{{\bm\varphi} \in K} {\cal S}({\bm\varphi}), \,\, \forall
K\subset {\cal H}.
\end{eqnarray}

Note that even though the LDP and the G-E theorem are formulated above for a discrete set
of random variables,
%*
%${\bm\varphi}_{t}$ parameterized by $t\in\mathbb{N}$,
the formulation also
implies straightforward extension to the continuous parameterization
$t\in\mathbb{R}_{+}=\{\tau\in \mathbb{R}|\tau>0\}$, where $t$ plays the role of time.
In particular, we choose
${\bm\varphi}_{t}=(\rho_{t},{\bm J}_{t})\in {\cal H}$ with ${\rm div}{\bm J}_{t}=0$;
the empirical density and current density are introduced in
Eqs.~(\ref{rho-jay1}) and (\ref{rho-jay2}), respectively.
%generated by closed stochastic trajectories ${\bm\eta}:S^{1}\to M$ of duration $t$.
In accordance with
discussion of Section \ref{sec:currents-cycles}, the argument ${\bm\psi}=(V,{\bm A})$ of the
generating function is represented by a potential function $V$ conjugate to the density, and by a
gauge equivalence class of an Abelian zero curvature gauge field ${\bm A}$ conjugate to the
current density.
The LDP can be viewed as a mathematically correct way to
formulate a physically intuitive statement that under the described conditions at long enough times
$t$ the probability distribution adopts an asymptotic form
\begin{eqnarray}
\label{LDP-physical}
P_{t}({\bm\varphi})\sim \exp(-t{\cal S}({\bm\varphi})).
\end{eqnarray}

By its formulation, specifically due to Eq.~(\ref{define-Q-s}), the LDP allows the rate functions
for reduced variables to be obtained via a variational principle. Let ${\cal H}'$ be the residence
vector space for the reduced variables ${\bm\varphi}'_{t}$ with the reduction (projection) map
$p:{\cal H}\to {\cal H}'$. In our case the reduced variables are the currents
${\bm\varphi}'_{t}={\bm\omega}_{t}$, so that ${\cal H}'=H_{1}(M;\mathbb{R})$ and the reduction map,
obviously defined by $p(\rho_{t},{\bm J}_{t})=[{\bm J}_{t}]$, is linear. The conjugate to
${\bm\varphi}'_{t}$ argument ${\bm\psi}'$ of the generating function ${\cal Q}'({\bm\psi}')$ resides
in $({\cal H}')^{*}\cong H^{1}(M;\mathbb{R})$ and, therefore, can be represented ${\bm\psi}'=[{\bm
A}]$ by a gauge equivalence class of a curvature-free $d{\bm
A}=(1/2)(\partial_{i}A_{j}-\partial_{j}A_{i})dx^{i}\wedge dx^{j}=0$ vector field. Applying the LDP
we obtain
\begin{eqnarray}
\label{variational-gen} {\cal S}'({\bm\omega})={\cal S}(p^{-1}({\bm\omega}))=\inf_{[{\bm
J}]={\bm\omega}}{\cal S}(\rho,{\bm J}).
\end{eqnarray}

Moreover, the LDP can be further interpreted as effectively implementing Legendre-type
transformations between thermodynamic potentials (effective actions).  We describe these general
relations in the following paragraphs.

Let us also note that our stochastic theory of maps $S^{1}\times Y\to X$ is, in fact, a field
theory; however, being identified as a theory of stochastic trajectories $S^{1}\to M$, with $M={\rm
Map}(Y,X)$, it can be interpreted as classical one-dimensional statistical mechanics in a circular
system of the size $t$ with the target space $M$. The time-averaged current
${\bm\omega}_{t}=t^{-1}{\bm Q}_{t}$ can be viewed as the density of the topological charge ${\bm
Q}_{t}=[\delta S_{t}({\bm A})/\delta{\bm A}]$, obtained from the variational derivative of the
gauge-invariant action $S_{t}({\bm A})$ with respect to the stationary vector potential ${\bm A}$, followed by switching to
(homology) equivalence classes, the latter operation denoted by square brackets. Naturally, a gauge equivalence class $[{\bm A}]\in
H^{1}(M;\mathbb{R})$ of curvature free vector field ${\bm A}$ can be interpreted as the chemical
potential that corresponds to the topological charge ${\bm Q}$. According to Eqs. (\ref{fenergy}),
(\ref{LDP-physical})  the rate ${\cal
S}({\bm\omega})=t^{-1}\Omega_{t}({\bm\omega})$ and the logarithmic generation ${\lambda}({\bm
A})=t^{-1}W_{t}([{\bm A}])$ functions can be interpreted as the densities of the free energy and
thermodynamic potential, respectively, in the thermodynamic limit $t\to \infty$. Naturally, they
are connected via the Legendre transformation
\begin{eqnarray}
\label{Legendre-topological} d\Omega={\cal S}dt+[{\bm A}]\cdot d{\bm Q}, \;\;\; dW={\cal S}dt-{\bm
Q}\cdot d[{\bm A}],
\end{eqnarray}
and the thermodynamic potential is represented by the effective action:
\begin{eqnarray}
\label{define-W} e^{-W_{t}({\bm A})} = \left\langle e^{t\int_{M}{\bm J}_{t}\cdot {\bm
A}}\right\rangle =\int {\cal D}{\bm\eta}e^{-S({\bm\eta})+t\int_{M}{\bm J}_{t}\cdot {\bm A}},
\end{eqnarray}

Therefore, applying the LDP argument to the
%time-averaged
empirical current ${\bm\omega}_{t}$ will lead to a
rate function ${\cal S}$, which is simply the generation rate of the effective action for the
currents of the theory. This restatement of the problem would be rather trivial, unless the theory
had interesting topological structure.
Indeed, as known from quantum field theories \cite{Abdalla}, in such cases
currents may have non-perturbative, anomalous terms arising from global topological effects. As we
will show in the remainder of the paper, this also happens in the non-equilibrium stochastic theory
with driving.

\subsection{Path-integral picture of LDP and the current density functional}
\label{subsec:LDP-path-integral}

In this Subsection we, first, derive the general (compact spaces) LDP expression
Eq.~(\ref{CDDF-continuous-explcit}) for the distribution of density current and density, and,
second, discuss respective transformation to the conjugated variables (potentials).

\subsubsection{Derivation of Eq.~(\ref{CDDF-continuous-explcit})}

Using a standard representation for the Dirac
$\delta$-functional in Eq. (\ref{prob}) gives for the probability distribution function
 \begin{align}
 \label{A-V-integral}
 &
\mathcal{P}_{t}(\bm{J}, \rho)\sim
 \int\!\!{\cal D}{\bm A}{\cal D}V
e^{-it\int d{\bm x}({\bm A}\cdot {\bm J}+V\rho)}\!\!
 \int\!\! {\cal D}{\bm{\eta}}_{t} e^{-S({\bm\eta}_{t};{\bm A},V)}, \\
 \label{S-modified}
 & S({\bm\eta}_{t};{\bm A},V)=S({\bm\eta}_{t})-i \int_{0}^{t} d \tau
 \left( \dot{\eta}_{t}^{j}A_{j}({\bm\eta}_{t})+ V({\bm\eta}_{t})\right),
 \end{align}
where the integration is performed over real auxiliary fields ${\bm A}$ and $V$,
and ${\cal D}{\bm A}$ and ${\cal D}V$ are the standard field-theoretical notations for functional
differentials/measures.

In the large deviation limit ($t\to\infty$),  the path
integral in (\ref{S-modified}) is estimated as
\begin{align}
\label{FAV}
& \int\!\! {\cal D}{\bm\eta}_{t} e^{-S({\bm\eta}_{t};{\bm A},V)}
={\rm Tr}e^{t\hat{{\cal L}}_{{\bm A},V}}
\sim\exp\left(t\lambda(\bm{A}, V)\right),
\end{align}
%\end{eqnarray}
with $-\lambda(\bm{A}, V)$ being the lowest eigenvalue of the operator $-\hat{{\cal L}}_{{\bm A},V}$,
\begin{align}
\hat{{\cal L}}_{{\bm A},V}\bar{\rho}=\lambda\bar{\rho},
\mbox{ where} \quad
\hat{{\cal L}}_{{\bm A},V}=(\kappa/2)\nabla_j\nabla_j - \nabla_{j}F_{j}+iV,
\quad  \nabla_j \equiv\partial_j - i A_j.
\label{FP-modified}
\end{align}
For the normalized ground state eigenfunction $\bar{\rho}({\bm x})$ we obtain
\begin{align}
\label{lambda}
& \lambda(\bm{A}, V)=\int d{\bm x}
\hat{{\cal L}}_{{\bm A},V}\bar{\rho}({\bm x})
\quad\mbox{and}\quad
\int d{\bm x}\bar{\rho}({\bm x})=1\,.
\end{align}

Applying further the saddle-point approximation to the functional integral in
Eq. (\ref{A-V-integral}) with respect to ${\bm A},V$, and using
Eqs. (\ref{S-modified})--(\ref{lambda}), we arrive at the following equations:
\begin{eqnarray}
 {\bm J}(x)=({\bm F}+i\kappa{\bm A}-(\kappa/2)\bm\partial)\bar{\rho}({\bm x}),\ \ \bar{\rho}=\rho.
 \label{sp}
\end{eqnarray}
%*
Actually, this saddle-point approximation involves a deformation of the integration contours
to the complex plane, which makes $\bm A$ imaginary and $V=0$ in the saddle point.
Thus, the saddle-point approximation corresponds to the supremum with respect to $(-i{\bm A})$
in the G-E theorem (in subsection \ref{subsec:LDP-review} the definition of $\bm A$ differs from
that of this subsection by a factor $i$).
Solving Eqs. (\ref{lambda}, \ref{sp}) for ${\bm A},V$ and $\bar{\rho}$ and substituting the result back
in the saddle-point expression for the integral in
Eq. (\ref{A-V-integral}) yields Eq. (\ref{CDDF-continuous-explcit}) for the Cram\'er functional.
Note that boundary (surface) terms do not contribute due to the compactness of the
target and base manifolds. In deriving Eq. (\ref{FAV}) from Eq. (\ref{S-modified}), the gauge
freedom was fixed by the requirement that the left eigenfunction of $\hat{\cal L}_{{\bm A},V}$,
conjugated to the right eigenfunction $\bar{\rho}$, equals unity.

\subsubsection{LDP for charges generated by scalar and vector potentials}

The explicit large deviation result (\ref{CDDF-continuous-explcit}) can be immediately used to get
thermodynamics-like relations for Cram\'er functions of derived objects. One introduces the sets
$\bm{A}^{(a)}({\bm x})$ and $V^{(b)}({\bm x})$ of vector and scalar potentials, respectively (which
can also be interpreted as gauge fields, i.e. generators of continuous symmetry transformations
mentioned above), and the corresponding sets $\bm{w}_{{\bm A}}({\bm J}_{t})$ and
$\bm{u}_{V}(\rho_{t})$ of charges
\begin{eqnarray}
 \label{define-J-rho} w_{{\bm A}}^{(a)}({\bm J}_{t})&\equiv&\int_{M}d{\bm x}A_{j}^{(a)}({\bm x})
 J_{t}^{j}({\bm\eta};{\bm
x})=t^{-1}\int_{0}^{t}d\tau \dot{\eta}^{j}A_{j}^{(a)}({\bm \eta}), \nonumber \\
u_{V}^{(b)}(\rho_{t})&\equiv&\int_{M}d{\bm x}V^{(b)}({\bm x})\rho_{t}({\bm\eta};{\bm
x})=t^{-1}\int_{0}^{t}d\tau V^{(b)}({\bm\eta}(\tau)).
\end{eqnarray}
At $t\to\infty$, the joint p.d.f. $\mathcal{P}_{t}(\bm{w},\bm{u})\equiv \langle
\delta(\bm{w}-\bm{w}_{{\bm A}}({\bm J}_{t}))\delta(\bm{u} - \bm{u}_{V}(\rho_{t}))\rangle_{\xi}$ of
$\bm{w}_{{\bm A}}({\bm J}_{t})$ and $\bm{u}_{V}(\rho_{t})$ has the large deviation form
 $\mathcal{P}_{{\bm A},V} (\bm{w},\bm{u})\sim \exp(-t{\cal S}_{{\bm A},V}(\bm{w},\bm{u}))$,
where
\begin{align}
\label{variational-pinciple} & {\cal S}_{{\bm A},V}(\bm{w},\bm{u})=\inf_{\bm{w}_{{\bm A}}({\bm
J})=\bm{w}, \; \bm{u}_{V}(\rho)=\bm{u}} {\cal S}( {\bm J},\rho).
\end{align}

In the path-integral terms, the variational principe expression (\ref{variational-pinciple}) can be
obtained by representing the probability distribution
\begin{align}
\label{derive-var-principle}  & \mathcal{P}_{{\bm A},V}({\bm w},{\bm u})\sim \int{\cal D}{\bm
J}{\cal D}\rho\delta({\bm w}-{\bm w}_{{\bm A}}({\bm J}))
\delta({\bm u}-{\bm
u}_{V}(\rho))e^{-t{\cal S}({\bm J},\rho)} \, ,
\end{align}
followed by computing the integral in the $t\to\infty$ limit using the saddle-point approximation.

Considering a marginalized version of Eq.~(\ref{variational-pinciple}), associated with the
distribution functions of the charges, generated by the vector potentials only, we have
$\mathcal{P}_{{\bm A}}(\bm{w})\sim \exp(-t{\cal S}_{{\bm A}}(\bm{w}))$ with
\begin{align}
\label{variational-pinciple-current} & {\cal S}_{{\bm A}}(\bm{w})=\inf_{\bm{w}_{{\bm A}}({\bm
J})=\bm{w}}{\cal S}( {\bm J},\rho).
\end{align}

The variational principle, represented by Eq.~(\ref{variational-pinciple-current}) has two
important implications that correspond to two specific choices of the gauge field sets ${\bm
A}^{(a)}({\bm x})$. The choice ${\bm A}({\bm x})={\bm F}({\bm x})$ leads to the observable $w_{{\bm
F}}({\bm J}_{t})$ that, obviously, represents the work (entropy) production rate, whose Cram\'er
function ${\cal S}(w_{{\bm F}})$ satisfies the fluctuation theorem.

However, in this manuscript we focus mainly on the other implication of
Eq.~(\ref{variational-pinciple-current}) associated with the set $\{{\bm A}^{(a)}\}$ of
curvature-free $d{\bm A}^{(a)}=0, \; \forall a$ is chosen in a way so that the corresponding set
$\{[{\bm A}^{(a)}]\}$ of equivalence classes forms a basis set in $H^{1}(M;\mathbb{R})$. According
to the Poincar\'e duality, as described at the end of section \ref{subsec:current-density-Poicare} and
specifically due to Eqs.~(\ref{Poincare-duality}) and (\ref{Poincare-duality-compliment}) the
current ${\bm\omega}=[{\bm J}]$ as the homology class of the current density ${\bm J}$ is fully
characterized by the values of the set $w_{{\bm A}^{(a)}}({\bm J})$ of observables. Therefore, for
the described choice of $\{{\bm A}^{(a)}\}$ the variational principle of
Eq.~(\ref{variational-pinciple-current}) is equivalent to the variational principle of
Eq.~(\ref{variational-gen}).

We should note for completeness that since $\bm{\omega}\in H_{1}(M;\mathbb{R})$, the Fourier variable
$\bm{\psi}$ conjugate to $\bm{\omega}$ resides in the space $\bm{\psi}\in
(H_{1}(M;\mathbb{R}))^{*}\cong H^{1}(M;\mathbb{R})$,
which results in the following representation:
\begin{align}
\label{prob-omega} \mathcal{P}_{t}({\bm\omega})\equiv \langle \delta({\bm\omega}_{t}-
{\bm\omega})\rangle_{\xi}\sim
\int_{H^{1}(M;\mathbb{R})}d{\bm\psi}e^{it{\bm\psi}\cdot({\bm\omega}-{\bm\omega}_{t})}
\,.
\end{align}
Going along the lines of derivation of
Eq.~(\ref{S-modified}) and using the Poincar\'e duality (\ref{Poincare-duality-compliment}),
which allows the identification ${\bm\omega}=[{\bm J}]$ and ${\bm\psi}=[{\bm A}]$, we recast
Eq.~(\ref{prob-omega}) as
\begin{align}
\label{prob-omega-2} & \mathcal{P}_{t}([{\bm J}])\sim \int_{H^{1}(M;\mathbb{R})}d[{\bm
A}]e^{-it\int_{M}d{\bm x}{\bm A}\cdot{\bm J}}\int{\cal D}{\bm\eta}_{t}e^{-S({\bm\eta};{\bm A},0)}
\nonumber \\ & \sim \int_{H^{1}(M;\mathbb{R})}d[{\bm A}]e^{-it\int_{M}d{\bm x}{\bm A}\cdot{\bm
J}}e^{t{\lambda}({\bm A},0)} \, .
\end{align}
This representation implies that ${\lambda}([{\bm A}])={\lambda}({\bm A},0)$ is obtained from
${\lambda}({\bm A},V)$ by restricting the to the zero $V=0$ scalar potentials and curvature-free $d{\bm
A}=0$ vector potentials.

\subsection{Derivation of the Cram\'er functional for topological currents}
\label{subsec:general-topo-currents-functional-derivation}

This Subsection describes a general strategy for calculating the Cram\'er functional
${\cal S}(\bm\omega)$ of the topologically
protected currents $\bm\omega$.
This is achieved via the variational procedure, formulated in subsection~\ref{subsec:LDP-review}
[Eq.~(\ref{variational-gen})], equivalent to ``integrating out" current density ${\bm J}$ and
density $\rho$ dependence for a fixed value of ${\bm \omega}$.

Specifically,
we will minimize the Cram\'er functional (\ref{CDDF-continuous-explcit})
${\cal S}({\bm J}, \rho)$ over ${\bm J}$ and $\rho$ under the following conditions
\footnote{Throughout the paper we use three equivalent representations of the current density:
vector field, $1$-form and $(m-1)$-form. The vector field is related to the $1$-form
through the natural metric tensor that characterizes noise correlations.
The relation between the Hodge dual differential forms of degrees $1$ and $(m-1)$ is also determined
by the metric tensor.}:
%-------------------------------------------------------------------------------------
\begin{eqnarray}
\label{conditions}
{\rm div}{\bm J} \equiv d^{\dagger}{\bm J}=0\,,
\quad
\int d{\bm x} \rho = 1\,,
\quad
[\bm J] = \bm \omega
\,.
\end{eqnarray}
A variation of the current density that satisfies the continuity condition and keeps the topological
current constant has a form
$\delta{\bm J}=d^{\dagger}{\bm \zeta}$ with ${\bm \zeta}$ being a
$2$-form. A straightforward calculation allows the requirement $\delta {\cal S}/\delta{\bm\zeta}=0$
to be represented in a form:
%-------------------------------------------------------------------------------------
\begin{eqnarray}
\label{var-eq-1} d{\bm A} \equiv {\bm \partial}\times{\bm A}=0\quad \mbox{with} \quad {\bm A}=\rho^{-1}\left({\bm
F}\rho-{\bm J}-(\kappa/2){\bm \partial}{\rho}\right),
\end{eqnarray}
where the second equality in Eq.~(\ref{var-eq-1}) should be viewed as the definition of the vector
field ($1$-form) $\bm A$. The stationary current that minimizes ${\cal
S}({\bm\omega})$ corresponds to ${\bm A}=0$.
The vector potential $\bm A$ determines how the density and current density distributions locally differ at the given
topological current and in the stationary regime.  Note that the vector field introduced in
Eq.~(\ref{var-eq-1}) differs from its counterpart introduced earlier in Eq.~(\ref{sp}) by a factor
$(-i\kappa)$. Thus, with a minimal abuse, we use the same notation for both and hereafter stick to
the one given by Eq.~(\ref{var-eq-1}).
%-------------------------------------------------------------------------------------

Variation of ${\cal S}({\bm J}, \rho)$ with respect to $\rho$ is performed in a straightforward way
by introducing a Lagrangian multiplier $\lambda$ to satisfy the normalization condition [the second
relation in Eq.~(\ref{conditions})]. This results in
%-------------------------------------------------------------------------------------
\begin{eqnarray}
\label{var-eq-2}
(\kappa/2){\rm div}\bm A+\bm F \cdot \bm A-(1/2){\bm A}^{2}=-\kappa\lambda\,.
\end{eqnarray}
%-------------------------------------------------------------------------------------
Representing ${\bm A}$ as a gradient
\begin{eqnarray}
{\bm A}=-\kappa{\bm \partial}\ln\rho_{-},
\label{rho-}
\end{eqnarray}
where $\rho_-(x)$ is a newly introduced scalar function, and substituting Eq.~(\ref{rho-})
into Eq.~(\ref{var-eq-2}), one finds that the quadratic first-order Riccati-type equation
(\ref{var-eq-2}) is transformed into the following linear second-order differential equation
\begin{eqnarray}
\label{var-eq-lin-1}
{\cal L}^{\dagger}\rho_{-}({\bm x})=\lambda\rho_{-}({\bm x})
\end{eqnarray}
%-------------------------------------------------------------------------------------
with the adjoint Fokker-Planck operator
%-------------------------------------------------------------------------------------
\begin{eqnarray}
\label{var-eq-lin-2}
{\cal L}^{\dagger} = (\kappa/2){\bm \partial}^2 + \bm F \cdot \bm \partial\,.
\end{eqnarray}
%-------------------------------------------------------------------------------------
On a compact manifold $M$, eigenvalues $\lambda$ are discrete and bounded from below. A solution
of Eq. (\ref{var-eq-lin-1}) corresponding to the lowest $-\lambda$ determines $\rho_-(x)$ and
$\lambda$. However, the function $\rho_{-}(x)$ is not necessarily single-valued and can acquire
uncertainty in the result of going around topologically nontrivial cycles.
We can find a unique solution ${\bm A}$ of Eq.~(\ref{var-eq-lin-2}) with the minimal eigenvalue if
we fix a set $\bm Z=(Z_{k}|k=1,\cdots,n)$ of topological parameters defined by the integrals
\begin{eqnarray}
\label{Zi}
\ln Z_{k}=\kappa^{-1}\int_{s_{k}}A_{i}({\bm x})dx^{i}
\end{eqnarray}
over the set $\{s_{k}\}_{k=1,\ldots,n}$ of the topologically independent $1$-cycles of $M$.

To summarize, Eqs.~(\ref{conditions})-(\ref{var-eq-2}) constitute a complete set of equations that
can be used to obtain the Cram\'er functional ${\cal S}({\omega})$ as well as the distributions
$\rho$ and ${\bm J}$. This is achieved in three steps.

\begin{itemize}
\item [(1)] We first solve Eq.~(\ref{var-eq-2})
together with the constraint $d{\bm A}=0$ (i.e., the first relation in Eq.~(\ref{var-eq-1})) with
respect to the vector potential ${\bm A}$. As explained above this is equivalent to solving the
linear lowest eigen-value problem (\ref{var-eq-lin-1}) with the additional topological freedom
fixed unambiguously selecting the sets $\bm Z=(Z_{k}|k=1,\cdots,n)$.

\item [(2)] We combine the first two relations in Eq.~(\ref{conditions}) with the second relation in
Eq.~(\ref{var-eq-1}), which results in
\begin{eqnarray}
\label{eq-generic-J-rho} (\kappa/2){\bm\partial}\rho-({\bm F}-{\bm A})\rho=-{\bm J}, \quad
d^{\dagger}{\bm J}=0,
\quad
\mbox{and}
\quad
\int d{\bm x} \rho = 1.
\end{eqnarray}
>From this linear system we find $\rho$ and $\bm J$ in terms of $\bm Z$ parameterizing ${\bm A}$
\footnote{ This derivation is similar to the one given in Ref. \cite{LandauerSwanson}
for the stationary current caused by a force field; in our case the force field is replaced by $\bm F - \bm A$.}.

\item [(3)] The Cram\'er functional ${\cal S}$ can be obtained as a function of ${\bm Z}$ by
substituting the solution obtained on steps (1) and (2) into Eq.~(\ref{CDDF-continuous-explcit}).
Finally, the substitution of the obtained solution for ${\bm J}$ into the third relation in
Eq.~(\ref{conditions}) establishes a relation between ${\bm Z}$ and $\bm\omega$, which, being
resolved with respect to ${\bm Z}$ in terms of ${\bm\omega}$, results in ${\cal S}({\bm\omega})$.
\end{itemize}

In the case $d{\bm F}=0$ of topological driving, the system of equations
Eqs.~(\ref{conditions})-(\ref{var-eq-2}) for the Cram\'er functional can be further simplified. One
applies the operator $(\kappa/2)d-({\bm F}-{\bm A})\wedge$ to the second relation in
Eq.~(\ref{var-eq-1}) and makes use of the relations $d^{2}=0$, $d{\bm A}=0$, and $d{\bm F}=0$.
This results in $((\kappa/2)d-({\bm F}-{\bm A})\wedge){\bm J}=0$. Combined with the generic
expression for the Cram\'er functional (\ref{CDDF-continuous-explcit}) and applying some reordering
of terms, this allows the system of equations (\ref{conditions})-(\ref{var-eq-2}) and
Eq.~(\ref{CDDF-continuous-explcit}) to be recast as
\begin{align}
\label{S-omega-eq-1}
& (\kappa/2){\rm div}\bm A+\bm F \cdot \bm A-(1/2){\bm A}^{2}=-\kappa\lambda,
\;\;\; \kappa^{-1}\int_{s_{k}}A_{i}({\bm x})dx^{i}=\ln Z_{k}, \;\;\; d{\bm A}=0,
 \\
\label{S-omega-eq-2}
& (\kappa/2)d{\bm J}-{\bm F}\wedge{\bm J}+{\bm A}\wedge{\bm J}=0, \;\;\;
d^{\dagger}{\bm J}=0, \\
\label{S-omega-eq-3}
& (\kappa/2)d\rho-{\bm F}\rho+{\bm A}\rho=-{\bm J}, \int_{M}d{\bm x}\rho({\bm
x})=1, \\
\label{S-omega-eq-4}
& {\cal S}({\bm\omega})=(2\kappa)^{-1}\int_{M}d{\bm x}\rho{\bm A}^{2}, \;\;\;
\int_{\alpha_{k}}{\bm J}=\omega_{k},
\end{align}
where for $k=1,\ldots,n$ one denotes by $s_{k}$ and $\alpha_{k}$ the dual sets of topologically
independent $1$- and $(m-1)$-cycles, respectively, i.e., $[s_{k}]$ and $[\alpha_{k}]$ form the basis
sets of $H_{1}(M)$ and $H_{m-1}(M)$, with the intersection property
$s_{k}*\alpha_{k'}=\delta_{kk'}$. Note that in Eqs. (\ref{S-omega-eq-2}) the current density
is understood as the $1$-form.

Therefore, the procedure of finding ${\cal S}({\bm\omega})$ in the case of the topological driving
$d{\bm F}=0$ can be summarized as follows:
\begin{itemize}

\item [(i)] For an arbitrary set ${\bm Z}=(Z_{k}|k=1,\ldots,n)$ we find a unique solution of
Eq.~(\ref{S-omega-eq-1}) that corresponds to the minimal value of $\lambda$ and express the vector
potential ${\bm A}$ in terms of the parameter set ${\bm Z}$.

\item [(ii)] Upon substitution of ${\bm A}$ found on the first step into Eq.~(\ref{S-omega-eq-2}), the
latter can be viewed as a system of homogeneous linear equations for the current density ${\bm J}$,
whose solution is unique up to a multiplicative factor.

\item [(iii)] We substitute the obtained current density ${\bm J}$ and vector potential ${\bm A}$
into Eq.~(\ref{S-omega-eq-3}), which is now viewed as a linear equation on the density $\rho$ with
an inhomogeneous term represented by ${\bm J}$.
Therefore, the normalization condition (the second equality in
Eq.~(\ref{S-omega-eq-3})) determines the prefactor in the current density, which identifies
normalization (multiplicative) factors both for ${\bm J}$ and ${\rho}$.

\item [(iv)] We substitute ${\bm A}$, ${\bm J}$, and $\rho$ into Eq.~(\ref{S-omega-eq-4}). This
expresses the Cram\'er function ${\cal S}$ and the topological current ${\bm\omega}$ in terms of
${\bm Z}$. Expressing ${\bm Z}$ in terms of ${\bm\omega}$ gives the desired Cram\'er function
${\cal S}({\bm\omega})$.
\end{itemize}

It is also instructive to note that an alternative representation for the Cram\'er functions
%-------------------------------------------------------------------------------------
\begin{eqnarray}
\label{functional-simplified} {\cal S} = -\lambda-\kappa^{-1}\int d{\bm x} \, \bm J\cdot\bm A =
(2\kappa)^{-1}\int d{\bm x} \rho \bm A^2\,,
\end{eqnarray}
%-------------------------------------------------------------------------------------
allows the eigenvalue $\lambda$ to be determined, which is useful for the weak-noise calculation.

The formal scheme described above will be implemented explicitly on our enabling example of the
topologically driven spin chain in Section \ref{subsec:derive-Cramer}.

\section{Non-Equilibrium Cyclic Spin-Chain}
\label{sec:spin-chain}

This section focuses on the applications of the general formalism developed above to the model of
topologically driven system of $N$ classical spins, arranged in a circular chain. In the
thermodynamic, $N\to \infty$, limit the configuration space becomes infinite-dimensional, and the
system can be viewed as a $(1+1)$ stochastic field theory with
the target space
$S^{2}$ (see Fig.~\ref{stringchain}). The model has three parameters $v$, $u$, and $\kappa$
that describe the relaxation rate, the rate of topological driving and the noise strength,
respectively. As briefly discussed above, the relaxation term suppresses the short-range
fluctuations, and, therefore, makes the model divergence-free in the thermodynamic limit. We are
considering the weak-noise limit, $\kappa \ll v,|u|$, whereas $v$ and $|u|$ can be comparable
although such that $|u|/v$ is not too large. The last requirement translates into the condition that
the constant loop ${\bm n}(y)={\bm n}_{0}$ solution  constitutes a stable stationary point of the
deterministic (zero noise) dynamics.

This Section is organized as follows. Subsection \ref{subsec:SpinChainTop} is devoted to
formulation of the topological driving in terms of a multi-valued potential, referred to as the
Wess-Zumino potential. We also briefly discuss the finite-dimensional (regularized) approximations
for the infinite-dimensional field-theory configuration. Some details on the spin-chain
regularizations are presented in Appendix~\ref{app:top-config-space}, where we argue that, starting
with a large enough $N$, the relevant topological properties of the finite-dimensional approximations
stabilize to their continuous limit counterpart. In Subsection~\ref{subsect:LDP-instantons} we
describe the instanton (optimal fluctuation) mechanism of the current generation and identify the
structure of the instanton space $M_{0}\subset M$ that consists of all the configurations the
optimal trajectories (instantons) pass through. Subsection \ref{subsec:derive-Cramer} contains the
derivation of the main results concerning the spin-chain model. Here, we calculate the Cram\'er
function of the topological current $\omega$, generated in the system, by implementing the general
procedure outlined in Section \ref{subsec:general-topo-currents-functional-derivation}.
This is achieved by
solving the relevant Fokker-Planck type equations equations, using an ansatz for the density
$\rho$, current density ${\bm J}$, and the vector potential ${\bm A}$. The implemented ansatz is
asymptotically exact in the weak-noise limit. The Cram\'er function ${\cal S}(\omega)$ is derived
for $|\omega|\ll |v|(\ln(|v|/\kappa))^{-1}$, i.e., in the instanton gas regime.

\subsection{Wess-Zumino interpretation of the Circular Spin Chain Model}
\label{subsec:SpinChainTop}

Generally, a system described by Eq.~(\ref{Ito-eq}) is not globally driven if the force field
is given by an exact differential, i.e., ${\bm F}=-dV$ with $(dV)_{i}=\partial_{i}V$, where $V$ is
some scalar potential function.
Although the force field form is not exact in the spin-chain model, it
is still {\emph{closed}}: $d{\bm F}=0$, with
$(dF)_{ij}=(1/2)(\partial_{i}F_{j}-\partial_{j}F_{i})$, or, in other words,
the force has \emph{zero curvature}.
Therefore, the deterministic force in Eq. (\ref{eq-determ-explicit})
can be expressed as
\begin{eqnarray}
\label{potForm} && \bm F =-\left.\frac{\delta}{\delta {\bm n}} V_{WZ}({\bm n}) \right|_{{\bm
n}^2=1}
\end{eqnarray}
in terms of a multi-valued potential $V_{WZ}({\bm n})$. Without a single-valued potential, the
system is globally driven, and its stationary state can only be a non-equilibrium one, with a
current being generated.

To determine the potential, consider some reference configuration ${\bm n}_{0}\in M$, e.g., a
constant loop (a set of collinear spins) ${\bm n}_{0}(y)={\bm n}_{0}$
and for an arbitrary configuration ${\bm n}\in M$ represented by ${\bm n}(y)$ consider a path
${\bm\chi}:[0,t]\to M$, represented by ${\bm\chi}(y,s)$ with $|{\bm\chi}(y,s)|=1$ along the path
that connects ${\bm n}_{0}$ to ${\bm n}$,
i.e., ${\bm\chi}(0)={\bm n}_{0}$ and ${\bm\chi}(t)={\bm n}$, or, equivalently,
${\bm\chi}(y,0)={\bm n}_{0}$ and ${\bm\chi}(y,t)={\bm n}(y)$.
In the following we will skip the dependence on $(y,\tau)$ when obvious.
To derive the potential, we can simply integrate the force $\bm F$ along the path $\bm\chi$.
According to the Stokes theorem,
$d{\bm F}=0$ guarantees that we obtain the same potential
if the paths ${\bm\chi}$ and ${\bm\chi}'$ used in the integration are topologically equivalent.
Thus, we obtain the potential
%-------------------------------------------------------------------------------------
\begin{eqnarray}
\label{WZ-potential} \ && V_{WZ}({\bm n})=V_{\rm e}({\bm n})-u\varphi_{B}({\bm n})
%-\int_{{\bm\chi}}F_{j}dx^{j}
, \quad V_{\rm e}({\bm n})=\frac{v}{2}\int_{S^1} dy (\partial_y {\bm n})^2,\\
&&
\varphi_{B}({\bm n})=-\int_{0}^{t}d\tau\int_{S^1} dy ({\bm\chi}\cdot[\partial_\tau
{\bm\chi},\partial_{y}{\bm\chi})] %=-i\ln r(\theta({\bm n}))
\,.\label{top_term}
\end{eqnarray}
%-------------------------------------------------------------------------------------
The ``elastic'' globally potential term $V_{\rm e}({\bm n})$, whose variation produces
${\bm F}_{\rm e}({\bm n})$ in Eq.~(\ref{eq-determ-explicit}),
enforces relaxation of the spin system to
a $y$-uniform distribution (a constant loop).
The second term $\propto\varphi_B$
is topological and similar to the multi-valued Wess-Zumino action \cite{Abdalla}. To avoid confusion we note the in this context we are not talking about a Wess-Zumino term in the action of our (1+1) field theory that would have originated from integration of a closed $3$-form over a relevant $3$-cycle. We rather interpret the potential in Eq.~(\ref{WZ-potential}) as the action of a free particle, represented by $V_{e}$ (with $y$ playing the role of time), with an additional multi-valued term $\varphi_{B}$, obtained via integration of a closed $2$-form over a relevant $2$-cycle, as described by Eq.~(\ref{top_term}).
The quantity $\varphi_{B}({\bm n})$ is the area enclosed by the loop ${\bm n}(y)$.
More precisely, values $\varphi_{B}({\bm n})$ calculated with help of two paths
${\bm\chi}(\bm y, \bm s)$ and ${\bm\chi}'(\bm y, \bm s)$ can differ by $4\pi m$
with integer $m$ if the paths belong to different equivalence classes.
The notation $\varphi_{B}$ indicates that it is
the Berry phase \cite{Berry} associated with the
loop ${\bm n}(y)$ on a sphere.
Indeed, the parallel transport of a tangent vector on the unit sphere along the loop
${\bm n}(y)$ rotates the vector by the angle equal to the area enclosed by ${\bm n}(y)$.
The Berry phase appears in the related quantum-mechanical phenomenon:
the state of the spin with the projection $s$ along the magnetic field
acquires a phase factor $e^{is\varphi_{B}}$ when the direction of the slowly changing
magnetic field makes one turn of the loop $\bm n(\bm y)$.

Finally we note that the purely topological nature of driving, i.e., $d{\bm F}=0$, in the continuous
field-theory limit of the considered model can be violated by a regularization, represented by a
(finite) $2N$-dimensional system of $N$ classical spins, arranged in a circular chain. Therefore,
there is a question of how (and if at all possible) to regularize the second term in the expression for the
driving force Eq.~(\ref{eq-determ-explicit}) to preserve the topological nature of driving on the
regularized level. This can be achieved by implementing an approach based on
Eq.~(\ref{WZ-potential}). To define the topological term we note that in the
thermodynamic limit $N\to \infty$,  the values of the neighboring spins are close,
i.e., $|\bm{n}_{j+1}-\bm{n}_{j}|\ll 1$ and, therefore, we can uniquely connect $\bm{n}_{j}$ to
$\bm{n}_{j+1}$ with geodesic lines. This results in a piece-wise smooth loop $\tilde{{\bm n}}(y)$
with the desired topological potential $u\varphi_{B}(\tilde{{\bm n}})$.
We provide some mathematical details of the regularization in Appendix \ref{app:top-config-space}

\subsection{Instantons and tubular neighborhoods}
\label{subsect:LDP-instantons}

This Subsection contains a preliminary discussion of our strategy in dealing with the problem of
weak noise. The approach consists in reducing our original model to an effective stochastic model
on a circle, where the new reduced variable is the re-parameterized Berry phase
$\theta({\bm x})\in S^1$:
\begin{eqnarray}
\varphi_B({\bm n})=2\pi(1-\cos(\theta/2))\,.
\label{param}
\end{eqnarray}

The representation in Eq. (\ref{param}) is possible and useful due to the instanton (optimal
fluctuation) character of the current generation in this weak noise limit considered here. Then,
a typical configuration is a closed loop on the sphere $S^2$  (also referred to, here and later on,
as a ``string") that is almost shrunk to a point performing a diffusive random walk over $S^2$
However, this typical diffusive meandering does not generate a current. Instead, the current is
generated by rare events naturally occurring along the instanton trajectories illustrated in
Fig.~\ref{stringchain}. The instanton trajectory is an optimal fluctuation as it has the highest
probability (i.e., minimizes the Onsager-Machlup action) among the trajectories resulting in the
transition. An instanton trajectory starts with a string, originally shrunk to a point, opening up
into a ``plane" circle configuration and then shrinking back to a point at the opposite end on the
sphere. Such a fast and rare process generates a full cycle in the space of the re-parameterized
Berry phase $\theta$.

The first step of our computational strategy, detailed in the next Subsection, consists in adopting
the instanton approximation to evaluation of the current density ${\bm J}$. The approximation
means that the non-equilibrium current density ${\bm J}$ obtained from Eqs.
(\ref{conditions})--(\ref{var-eq-2}) is concentrated in the narrow tubes
near the configurations passed by the optimal
fluctuation trajectories, i.e., the configurations ${\bm x}$ represented by ``plane'' circles
embedded into $S^2$. The concept of tubes where the current density is localized was used, for
example, in Refs. \cite{CaroliCaroliRouletGouyet80,LandauerSwanson,CDS08,MaierStein,DMRH,DLMcCS,04TNK}.

Let $M_{0}\subset M$ be the subspace of these
configurations. We refer to this as the space of instanton configurations. For example, all four
configurations shown in Fig.~\ref{stringchain} belong to the instanton space $M_{0}$.
The instanton space $M_{0}$ has a simple structure that allows parametrization for the
instanton as well as for a small neighborhood $U_{{\rm curr}}\supset M_{0}$ where the current is
generated. First of all with a minimal abuse of notation we denote by $\partial M_{0}\subset
M_{0}$ the set of configurations ${\bm x}\in M_{0}$ with $\theta({\bm x})=0$, i.e., constant loops
${\bm x}(y)={\bm n}$, parameterized by their positions ${\bm n}\in S^{2}$ on the target sphere.
Obviously, $\partial M_{0}\cong S^{2}$. Since $\partial M_{0}\subset M$ is a compact manifold
embedded into the configuration space, it has a standard small tubular neighborhood $U_{0}\supset
\partial M_{0}$, whose points $({\bm n},{\bm\xi})\in U_{0}$ are parameterized by the position
${\bm n}$ on the sphere and the set ${\bm\xi}$ of transverse variables. We further show that
$M_{0}\setminus
\partial M_{0}\cong SO(3)\times (0,2\pi)$ is an embedded $4$-dimensional non-compact manifold, thus having
a small tubular neighborhood $U\supset M_{0}\setminus \partial M_{0}$. This view suggests the
following reparameterization of a ``plane'' circle ${\bm x}:S^{1}\to S^{2}$. Let ${\bm e}_{3}$ be a
unit vector, orthogonal to the circle plane, with the direction determined by the loop orientation,
e.g., by $\partial_{y}{\bm x}(0)$. We denote by ${\bm e}_{1}$ a unit vector in the direction from
the circle center to its origin ${\bm x}(0)$, and set ${\bm e}_{2}=[{\bm e}_{3},{\bm e}_{1}]$. Then
the oriented orthonormal basis set $({\bm e}_{1},{\bm e}_{2},{\bm e}_{3})$ represents an element
$g\in SO(3)$ of the orthogonal group. The fourth coordinate of ${\bm x}$ is its Berry phase
$\theta(\bm x)\in (0,2\pi)$. The points of the corresponding tubular neighborhood are parameterized
by $(g,\theta,{\bm\zeta})$ with ${\bm\zeta}$ representing the set of transverse variables.
Obviously, $U_{{\rm curr}}=U_{0}\bigcup U$ covers the instanton space $M_{0}$.

The aforementioned coordinate representation in the tubular neighborhood $U$ has the following explicit representation
\begin{eqnarray}
\label{coord-U-explicit} {\bm n}(y)=\sin(\theta/2)\cos y{\bm e}_{1}-\sin(\theta/2)\sin y{\bm
e}_{2}+\cos(\theta/2){\bm e}_{3} + \sum_{j}\sum_{a}\zeta_{j}\psi_{j}^{a}(y;\theta){\bm e}_{a},
\end{eqnarray}
where ${\bm\zeta}=(\zeta_{j}|j=1,2,\ldots)$ is the transverse deviation expanded in the transverse
modes with the components $\psi_{j}^{a}(y)$ that generally is parametrically dependent on the Berry
phase $\theta$. For a regularized version $j=1,\ldots,2N-4$. We further note that since
$0<\theta<2\pi$ in the region $U$, this region does not contain nontrivial $1$-cycles.
Therefore, the multi-valued potential $V_{WZ}$ restricted to $U$ can be represented by a
single-valued function, which can be expanded in the transverse variables as
\begin{eqnarray}
\label{expand-V-WZ}
V_{WZ}({\bm x})=V_{0}(\theta)+W(\theta)({\bm\zeta}\otimes{\bm\zeta})/2 \quad
V_{0}(\theta)=(\pi/2)(-v\cos\theta+4u\cos(\theta/2)),
\end{eqnarray}
where $W(\theta)$ is represented by a symmetric matrix with the matrix elements $W_{ij}(\theta)$.

%%%%%%%%%%%%%%%%%%%%%%%%%%%%%%%%%%%%%%%%%%%%%%%%%%%%%%%%%%%%%%%%%%%%
\begin{figure}[htp]
  \begin{center}
      \includegraphics[width=0.8\textwidth]{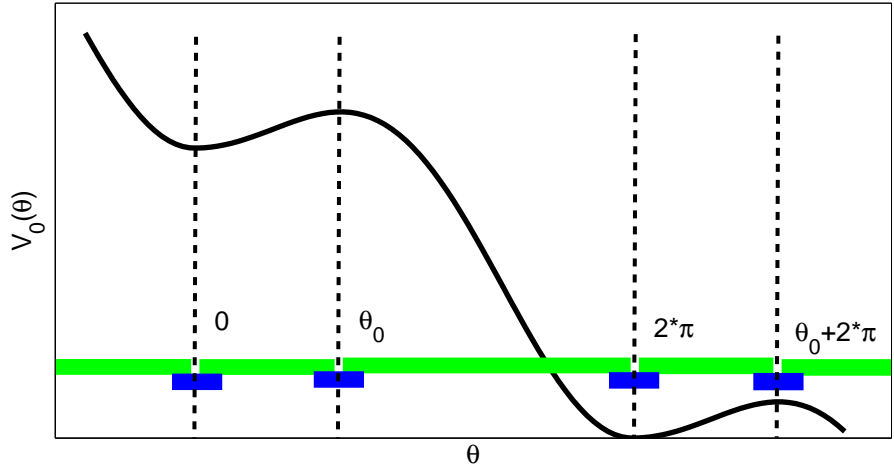}
  \end{center}
  \caption{Illustration of the $V_0(\theta)$ profile.
  Defined on the $[0,2\pi]$ span the potential is multi-valued.
  The bars show the overlapping harmonic (blue) and WKB (green) domains.}
  \label{fig:U0}
\end{figure}
%%%%%%%%%%%%%%%%%%%%%%%%%%%%%%%%%%%%%%%%%%%%%%%%%%%%%%%%%%%%%%%%%%%%%

The most important for us is the $\theta$-component of the force directed along the
instanton trajectory
%-------------------------------------------------------------------------------------
\begin{eqnarray}
\label{F0}
F_{0}(\theta)=-\partial_{\theta}V_{0}(\theta)=(\pi/2)(-v\sin\theta+u\sqrt{2(1-\cos\theta)})
\,.
\end{eqnarray}
%-------------------------------------------------------------------------------------
One observes that the effective potential $V_{0}(\theta)$ has two stationary points, $\theta=0$ and
$\theta=\theta_{0}$ where $F_{0}(0)=F_{0}(\theta_{0})=0$.
The unstable point
\begin{eqnarray}
\label{theta0}
\theta_{0}= 2\arccos(u/v)
\end{eqnarray}
is characterized by
\begin{eqnarray}
\label{k0}
k_{0}\equiv\partial_{\theta}F_{0}(\theta_{0})=(\pi/2) v\left(1-u^{2}/v^{2}\right)>0
\,.
\end{eqnarray}
The stable point $\theta=0$ can be approached along two directions corresponding to different
loop orientations. In terms of $\theta\in (0,2\pi)$ they correspond to $\theta\to +0$ and
$\theta\to 2\pi-0$; in both cases $\partial_{\theta}F_{0}<0$.
These two stationary points
correspond to saddle-point ($\theta=\theta_{0}$) and equilibrium ($\theta=0$) string configurations.
The main instanton approximation, valid in the low-noise
limit, means that only special instanton configurations, parameterized by $\theta$ and possibly
accounting for some small fluctuations (these from a small neighborhood $U\supset M_{0}\setminus
\partial M_{0}$), will be relevant to the discussion below.

We will make
some additional (to the basic instanton approximation) assumptions
limiting the domain of validity but also adding a required extra tractability in exchange.
In the following we will discuss separately: (a) The so-called WKB approximation that ignores the terms
associated with second-order derivatives over $\theta$ in comparison with the corresponding first-
and zero-order terms. The WKB approximation is valid in the WKB domain
where $\theta$ is sufficiently far away from the
equilibrium and saddle points: $\theta,2\pi-\theta \gg \sqrt{\kappa/k}$ and
$|\theta-\theta_{0}|\gg\sqrt{\kappa/k_{0}}$, respectively. The WKB region
naturally splits into two sub-domains, $0 \lesssim\theta\lesssim \theta_{0}$ and
$\theta_{0}\lesssim\theta\lesssim 2\pi$.
(b) The harmonic/ linear-force approximation is valid in
a relatively small vicinity of the saddle-point, $|\theta-\theta_0|\ll 1$, referred to as the
harmonic region, where one can use the linear approximation for the force, $F_{0}(\theta)\approx
k_{0}(\theta-\theta_{0})$. Similar to the WKB treatment in quantum mechanics, the WKB and harmonic
regions do overlap. See Fig.~\ref{fig:U0} for illustration. The concrete form of the instaton for the spin-chain model
in the harmonic
and WKB domains will be derived (and matched) in Section~\ref{subsec:derive-Cramer} and Appendix \ref{app:spin-chain-Cramer-detail}.
Finally we note that, as will be
demonstrated in Section~\ref{subsec:derive-Cramer}, detailed analysis in the harmonic region of
the stable point $\theta=0$ can be avoided.

\subsection{Derivation of the Cram\'er function}
\label{subsec:derive-Cramer}

In this Subsection we apply the strategy, outlined in Subsection
\ref{subsec:general-topo-currents-functional-derivation}, to the spin-chain model considered in the
weak-noise limit also requiring that the topological currents are not too strong. As outlined in
Subsection \ref{subsec:general-topo-currents-functional-derivation}, the vector potential
${\bm A}$ is characterized by the topological
parameters given by Eq.~(\ref{Zi}). As demonstrated in Appendix~\ref{app:top-config-space}, for
the spin-chain model $H_{1}(M)\cong\mathbb{Z}$, i.e., there is only one
topologically independent cycle $s$, which can be chosen to be restricted to the
instanton space $M_{0}$. The cycle $s$ for such a choice is the one illustrated in
Fig.~(\ref{stringchain}) and corresponds to just a change of $\theta$ from $0$ to $2\pi$. The dual
cycle $\alpha$ of codimension $1$ can be chosen to be determined by the condition $\theta({\bm
x})=\theta_{0}$, so that $s$ and $\alpha$ intersect at the saddle point. The topological current
for our model is single-component and will be denoted by $\omega$. The topological parameter is
also single-component,
%-------------------------------------------------------------------------------------
\begin{eqnarray}
\label{Z-A-relation} Z=\exp \left(\kappa^{-1}\int_{s}A_{i}({\bm x})dx^{i}\right).
\end{eqnarray}
%-------------------------------------------------------------------------------------

The four-step procedure for calculating ${\cal S}(\omega)$ in the relevant for our application case
$d{\bm F}=0$ of the topological driving, was outlined in Subsection
\ref{subsec:general-topo-currents-functional-derivation}. Implementation of this procedure to the
spin-chain model in the low-noise limit is facilitated by the following model-specific assumptions
for the ${\bm A}$, ${\bm J}$, and $\rho$ functions which can be verified directly once the
solution, based on these assumptions, is found:
\begin{itemize}
\item [(a)] The vector potential ${\bm A}$ is essentially nonzero only within the sub-domain
$|\theta-\theta_{0}|\lesssim \sqrt{\kappa/k_{0}}$, which is contained in the harmonic region of the
saddle point. In this domain ${\bm A}$ does not depend on the transverse variable ${\bm\zeta}$, and
its only substantially non-zero component, $A_{\theta}$, is along the $\theta$ variable.

\item [(b)] The eigenvalue $\lambda$ is exponentially small and can be neglected
in Eq.~(\ref{S-omega-eq-1}) everywhere except for a small vicinity of the equilibrium point
$\theta=0$. In this sub-domain  ${\bm A}$ is so small that it can be totally neglected. Moreover in
the entire WKB region ${\bm A}$ is still small enough, so that the nonlinear term ${\bm A}^{2}$ in
Eq.~(\ref{S-omega-eq-1}) can also be neglected.

\item [(c)] The main contribution to the integral for ${\cal S}(\omega)$ in Eq. (\ref{S-omega-eq-4})
comes from the $|\theta-\theta_{0}|\lesssim \sqrt{\kappa/k_{0}}$ domain.

\item [(d)] The current density ${\bm J}$ is concentrated in a small tubular neighborhood of
the instanton space $M_{0}$, whose transverse size scales $\sim\sqrt{\kappa}$ with the noise value. In the
WKB and saddle-point harmonic regions the longitudinal component of ${\bm J}$ has a Gaussian
dependence on ${\bm\zeta}$
\begin{eqnarray}
\label{J-theta-WKB}
J_{\theta}=J_{0}(\theta)e^{-\kappa^{-1}\sigma(\theta)({\bm\zeta}\otimes{\bm\zeta})}.
\end{eqnarray}
In the harmonic region of the saddle point this is the
only non-zero component of ${\bm J}$, and $\sigma(\theta)\approx W(\theta_{0})$.

\item [(e)] The density distribution $\rho$ is concentrated near the equilibrium $\theta=0$, where
rare events generating  the current can be neglected:
\begin{eqnarray}
\label{rho-harmonic-stable} \rho({\bm x}) \approx \rho_{0}e^{-2\kappa^{-1}V_{WZ}({\bm x})}.
\end{eqnarray}
\end{itemize}

The features of ${\bm A}$, ${\bm J}$, and $\rho$, listed above, are generic for a topologically
driven system in the low-noise limit.
$SO(3)$ symmetry is
an important special feature of our model.
Therefore, the critical points are represented by isolated orbits of $SO(3)$ rather than isolated
points. Discussing ``the'' equilibrium $\theta=0$ and ``the'' saddle
$\theta=\theta_{0}$ points in our model, what we actually mean is that we have two isolated orbits of $SO(3)$
which
represent the equilibrium and saddle points, respectively. These orbits are given by
$SO(3)/SO(2)\cong S^{2}$ and $SO(3)$, respectively. The symmetry gives rise to zero modes, which
substantially complicates a straightforward path-integral calculation in the instanton
approximation, especially due to the different numbers, $2$ and $3$, of zero modes at the
equilibrium $\theta=0$ an transition $\theta=\theta_{0}$ configurations, respectively.

In our approach, the zero modes are allowed for automatically in a very simple way in the
form of the volumes of the relevant orbits $S^{2}$ and $SO(3)$ for the equilibrium and saddle
configurations, respectively. \footnote{Our spin-chain model is degenerate and thus special, in what concerns
the number of stable points (only one) and number of zero modes at the stable and unstable fixed
points.  In a generic system the number of the equilibrium and saddle points can be arbitrary.
Moreover, the instanton manifold is $1$-dimensional. An instanton trajectory from the manifold
starts at a stable point ascends to a saddle point and consequently descends following an unstable
direction to another stable point. Therefore, $M_{0}$ in general can be viewed as a graph of
instanton transitions. Using the four-step strategy, outlined in
Subsection~\ref{subsec:general-topo-currents-functional-derivation}, the problem of finding the
Cram\'er function of $\omega$ can be reduced to a Markov chain model on the graph that represents
the instanton space $M_{0}$. This general approach, allowing to reduce a non-equilibrium
field-theory problem in the weak noise limit to a Markov chain model with states associated with
stable fixed points of the classical dynamics, will be addressed in a separate publication.}

Properties (a)-(e) listed above allow us to ease computation of the Cram\'er function ${\cal
S}(\omega)$ essentially and in particular bypass complications associated
with detailed resolution of the WKB domain. Therefore, in this Subsection we will rely on these
properties, presenting justification details related to WKB calculations, as well as some
technical details on the determinant calculations, in Appendix~\ref{app:spin-chain-Cramer-detail}
\footnote{Note also that the shape of the current density distribution ${\bm J}$ derived in
Appendix~\ref{app:spin-chain-Cramer-detail} provides with some useful information on the processes
that generate the current.}

The material in the remaining part of the Subsection is split into paragraphs according to the
four-step strategy described in
Subsection~\ref{subsec:general-topo-currents-functional-derivation}.

\subsubsection{Step (i): Identifying the vector potential}
\label{subsub:1}

In this Subsection we implement step (i). In the harmonic region $|\theta-\theta_{0}|\ll 1$,
according to property (a), the vector potential ${\bm A}$ is described by the only non-zero
component $A_{\theta}$ that depends on $\theta$ only. We seek a solution in the representation of
Eq.~(\ref{rho-}) with $\rho_{-}(\theta)$ depending on $\theta$ only. According to property (b) we
set $\lambda=0$ in Eq.~(\ref{var-eq-lin-1}). Making use of Eq.~(\ref{var-eq-lin-2}) we arrive at
the following homogeneous linear equation:
%-------------------------------------------------------------------------------------
\begin{eqnarray}
\label{rho-minus-saddle-eq}
(\kappa/2)\partial^2_{\theta}\rho_{-}+F_0(\theta)
\partial_{\theta}\rho_{-}=0\,,
\end{eqnarray}
%-------------------------------------------------------------------------------------
whose general solution can be expressed in terms of the error function ${\rm
erf}(z)=2\pi^{-1/2}\int_0^z ds \exp(-s^2)$:
%-------------------------------------------------------------------------------------
\begin{eqnarray}
\label{rho-minus-saddle-solution} \rho_{-}(\theta)/\rho_-(\theta_0) = 1+c{\,\rm erf}\left(
\sqrt{k_{0}/\kappa}(\theta-\theta_{0})\right)\,.
\end{eqnarray}
%-------------------------------------------------------------------------------------
This results in the following expression for the $\theta$-component of the vector potential:
%-------------------------------------------------------------------------------------
\begin{eqnarray}
\label{A-saddle-solution} A_{\theta} = -\sqrt{4\kappa k_{0}/\pi} \frac{c\exp\left(
-k_0(\theta-\theta_{0})^2/\kappa\right)} {1+c{\,\rm erf}\left(
\sqrt{k_{0}/\kappa}(\theta-\theta_{0})\right)}\,.
\end{eqnarray}
%-------------------------------------------------------------------------------------
To determine the constant $c$ we recast Eq.~(\ref{Z-A-relation}) as
$Z=\exp(\int_0^{2\pi}d\theta A_\theta /\kappa)$,  which results in
%-------------------------------------------------------------------------------------
\begin{eqnarray}
\label{c-Z-relation} c=(1-Z)/(1+Z) \,.
\end{eqnarray}
%-------------------------------------------------------------------------------------
Eq.~(\ref{A-saddle-solution}) shows that ${\bm A}$ peaks near $\theta_{0}$ and decays in a
Gaussian fashion as $\theta$ moves away from $\theta_{0}$. The solution presented in
Appendix~\ref{subapp:WKB-A} shows that the rapid decay continues with $\theta$ moving even further
away from $\theta_{0}$ inside the WKB domain.

\subsubsection{Steps (ii) and (iii): Identifying the current-density and density distributions}
\label{subsub:2}

The current density ${\bm J}$ does not have any special structure in the harmonic region
in contrast to the distributions ${\bm A}$ and
$\rho$. Therefore, on the step (ii),
the solution of Eqs. (\ref{S-omega-eq-2}) for ${\bm J}$ near the saddle point
can be extrapolated from the WKB region (see Appendix~\ref{subapp:WKB-J}).

Step (iii) starts with solving the first equation in Eqs.~(\ref{S-omega-eq-3}).
This is an easy task, since Eq.~(\ref{S-omega-eq-3}), being restricted to any $1$-dimensional
subspace, becomes a linear first-order ordinary differential equation with a right hand side.
On the cycle $s$ that belongs to the instanton space, the density $\rho$ is represented
by a function $\varrho(\theta)$, and the linear equation adopts a
form
%-------------------------------------------------------------------------------------
\begin{eqnarray}
\label{eq-rho-theta-reduced}
(\kappa/2)\partial_{\theta}\varrho(\theta)
-(F_{0}(\theta)-A_{\theta}(\theta))\varrho(\theta)=J_{\theta}(\theta),
\;\;\; \varrho(0)=\varrho(2\pi)=\rho_{0}
\end{eqnarray}
%-------------------------------------------------------------------------------------
The $1$-dimensional linear boundary problem, defined for $0 \le \theta \le 2\pi$ and represented by
Eq.~(\ref{eq-rho-theta-reduced}), can be solved in a standard way \cite{LandauerSwanson} by using a representation
%-------------------------------------------------------------------------------------
\begin{eqnarray}
\label{density-q} \varrho(\theta)= q(\theta) \exp\left(-2\kappa^{-1}\tilde{V}(\theta)\right), \quad
\tilde{V}(\theta) = V_{0}(\theta) +
\int_0^{\theta}A_{\theta}(\theta')d\theta'=\int_0^{\theta}(A_{\theta}(\theta')-F_{0}(\theta'))d\theta'
\end{eqnarray}
Substituting Eq.~(\ref{density-q}) into
Eq.~(\ref{eq-rho-theta-reduced}) one arrives at
%------------------------------------------------------------------------------------
\begin{eqnarray}
\label{density-q-eq} \partial_{\theta}q = -(2/\kappa)J_{\theta}e^{2\kappa^{-1}\tilde{V}(\theta)},
\;\;\; q(0)=\rho_{0}, \;\;\; q(2\pi)=\rho_{0}
\exp\left(2\kappa^{-1}\int_{0}^{2\pi}d\theta(A_{\theta}(\theta)-F_{0}(\theta))\right)
=\rho_{0}Z^{2}e^{S_{+}-S_{-}},
\end{eqnarray}
%-------------------------------------------------------------------------------------
where we have introduced the barriers
\begin{align}
\label{define-barriers} & S_{+} = -2\kappa^{-1}\int\limits_0^{\theta_0}d\theta\,F_0(\theta)\,,
\qquad S_{-} = 2\kappa^{-1}\int\limits_{\theta_0}^{2\pi} d\theta\,F_0(\theta).
\end{align}

Close to the saddle-point configuration, $q$ depends only on $\theta$,
and the only substantially nonzero component of the current density ${\bm J}$ is
%-------------------------------------------------------------------------------------
\begin{eqnarray}
\label{J-saddle-J0}
J_{\theta}({\bm x})= J_{0}e^{-\kappa^{-1}\sigma\bm\zeta\otimes\bm\zeta}\,,
\end{eqnarray}
%-------------------------------------------------------------------------------------
where $\sigma = W(\theta_0)$ and $J_{0}$ is constant on the length scale $\sqrt{\kappa/k_0}$,
according to the property (d).

Integration of Eq. (\ref{density-q-eq}) results in
%-------------------------------------------------------------------------------------
\begin{eqnarray}
\label{q-solution}
q(\theta)= \rho_0-\frac{2}{\kappa}J_0 e^{S_+}\int\limits_0^\theta d\tau
\exp\left( -k_0(\tau-\theta_0)^2/\kappa +(2/\kappa)\int\limits_0^\tau d\tau' A_\theta(\tau') \right)\,,
\end{eqnarray}
%-------------------------------------------------------------------------------------
which translates with the help of Eq. (\ref{A-saddle-solution}) into
%-------------------------------------------------------------------------------------
\begin{eqnarray}
\label{q-solution-explicit} q(\theta)= \rho_0-\sqrt{\pi/(\kappa k_{0})}J_0 e^{S_+}\frac{1-c}{c}
+\sqrt{\pi/(\kappa k_{0})}J_0 e^{S_+}
\frac{(1-c)^2}{c(1+c\,\mathrm{erf}(\sqrt{k_0/\kappa}(\theta-\theta_0)))} \,,
\end{eqnarray}
%-------------------------------------------------------------------------------------
where $c=(1-Z)/(1+Z)$. One observes from Eq.~(\ref{q-solution-explicit}) that $q(\theta)$ is
localized at $|\theta-\theta_{0}|\lesssim \sqrt{\kappa/k_0}$, thus justifying the approximations
used so far to evaluate $A_{\theta}$ and $J_{\theta}$. Setting $\theta=2\pi$ in Eq.
(\ref{q-solution-explicit}) and applying the boundary condition from Eq.~(\ref{density-q-eq}), one
expresses $J_0$ in terms of $Z$ and $\rho_{0}$:
%-------------------------------------------------------------------------------------
\begin{eqnarray}
\label{J0-Z} J_0= \rho_0\sqrt{\kappa k_{0}/(4\pi)}(Z^{-1} e^{-S_+} - Z e^{-S_-})\,.
\end{eqnarray}
%-------------------------------------------------------------------------------------

To complete the major step (iii) we apply the normalization condition, given by the second relation in
Eq.~(\ref{S-omega-eq-3}), to find the normalization constant $\rho_{0}$.
We consider the neighborhood  $U_{0}$
represented by the most probable configurations that are close to constant loops. Stated
differently, the density given by Eq.~(\ref{rho-harmonic-stable}) is concentrated in the harmonic
region of the stable point and hence adopts a form:
%-------------------------------------------------------------------------------------
\begin{eqnarray}
\label{rho-stable-result} \rho({\bm x})= \rho({\bm n},{\bm\xi})=
\rho_{0}e^{-\kappa^{-1}\gamma_{ij}\xi^{i}\xi^{j}}\,,
\end{eqnarray}
%-------------------------------------------------------------------------------------
where ${\bm n}\in S^2$ is the center of a small loop, and the matrix $\gamma$  determines a
harmonic expansion of $V_{WZ}(\bm x)$ in terms of nonzero modes ${\bm\xi}$.
Then the normalization condition becomes
%-------------------------------------------------------------------------------------
\begin{eqnarray}
\label{normalize-rho} \int_{S^2}d\mu({\bm n})\int{\cal D}{\bm\xi}\rho({\bm n},{\bm\xi})=1,
\end{eqnarray}
%-------------------------------------------------------------------------------------
thus leading to the following value of $\rho_{0}$ (see Appendix \ref{subapp:determinants} for more
details):
%-------------------------------------------------------------------------------------
\begin{eqnarray}
\label{normalize-multi-result} \rho_{0}=(1/4)\sqrt{\det\gamma}(\pi N)^{-N}\kappa^{1-N}.
\end{eqnarray}
%-------------------------------------------------------------------------------------
Note that due to the symmetry of the model $d\mu({\bm n})$ is an $SO(3)$-invariant measure on the
sphere, defined up to a multiplicative factor. This factor has been identified in a standard way by
considering the two zero modes on the sphere (see Appendix~\ref{subapp:determinants} for some
detail). The determinant $\det\gamma$ includes $(2N-2)$ positive eigenvalues of the discretized
operator $\gamma$.

\subsubsection{Step (iv): Finding the Cram\'er function for the topological current}
\label{subsub:3}

Step (iv) starts with evaluating the integral (\ref{S-omega-eq-4}) for the Cram\'er
function ${\cal S}$ in terms of $Z$. According to the property (d) the integral acquires its major
contribution from the vicinity of the saddle-point configuration. Indeed, the integrand decays as
$\propto \exp((2/\kappa)V_0(\theta))$ with $\theta$ deviating from $\theta_0$ inside the WKB
domains, where Eqs. (\ref{A-saddle-solution}), (\ref{c-Z-relation}), (\ref{density-q}) and
(\ref{q-solution-explicit}) can be used. The dependence of $\rho$ on the transverse variables in
the relevant region is given by $\rho(\bm x)=\varrho(\theta) \exp
(-\kappa^{-1}\sigma\bm\zeta\otimes\bm\zeta)$ with $\sigma=W(\theta_{0})$, which follows from
Eq.~(\ref{S-omega-eq-3}) applied in the transverse direction as well as from the asymptotic absence
of the transverse components of ${\bm A}$ and ${\bm J}$ in the relevant region.

Thus, we obtain the following expression for the Cram\'er function:
%-------------------------------------------------------------------------------------
\begin{eqnarray}
\label{functional-Arhodet} {\cal S}= (2\kappa)^{-1}\int_{SO(3)}d\mu_{\theta_{0}}(g) \int{\cal
D}{\bm\zeta}e^{-\kappa^{-1}\sigma\bm\zeta\otimes\bm\zeta} \int\limits_0^{2\pi} d\theta\,
A^2_\theta(\theta) \varrho(\theta)
\,.
\end{eqnarray}
%-------------------------------------------------------------------------------------
The integral over $\theta$ can be calculated similarly to that in
Eq. (\ref{q-solution})
whereas the calculation of the other integrals is presented in Appendix
\ref{subapp:determinants}:
%-------------------------------------------------------------------------------------
\begin{eqnarray}
\label{functional-J0} {\cal S}=\vartheta_{0}N^N(\pi\kappa)^{N-2}/{\sqrt{\det\sigma}} \left(
-J_0\ln Z+\sqrt{\kappa k_{0}/\pi}\rho_{0}(1-Z)(Ze^{-S_-}-e^{-S_+})/(2Z) \right)
\,.
\end{eqnarray}
%-------------------------------------------------------------------------------------
The integral over the space $SO(3)$ of the zero modes is computed explicitly:
%-------------------------------------------------------------------------------------
\begin{align}
\label{int-zero-modes-G}
\vartheta_{0}=\int_{SO(3)}d\mu_{\theta_{0}}(g)
=
2\pi^{2}(1+u^{2}/v^{2})\sqrt{1-u^{2}/v^{2}}\,,
\end{align}
%-------------------------------------------------------------------------------------
where the invariant measure $d\mu_{\theta_{0}}(g)$ in $SO(3)$, taking into account three zero
modes produced by infinitesimal changes of $g$, requires careful evaluation of the multiplicative
factor. This is performed in a standard way by considering the scalar products of these modes (see
Appendix \ref{subapp:determinants} for some detail).

To conclude the step (iv) we need to relate the topological current $\omega$ to the topological
parameter $Z$ by applying the second relation in Eq.~(\ref{S-omega-eq-4}). Let us reiterate that we
have chosen the cycle $\alpha$ by the condition $\theta({\bm x})=\theta_{0}$, so that we have for
the current
\begin{eqnarray}
\label{omega-J0}
\omega={\bm\omega}*[\alpha]= \int_{\alpha}{\bm J}=\int_{\theta({\bm
x})=\theta_{0}}d{\bm x}J_{\theta}(\bm x)\,,
\end{eqnarray}
which results in
\begin{align}
\label{expression-omega-G}
& \omega= J_{0}
\vartheta_{0}\int{\cal D}{\bm\zeta}e^{-\kappa^{-1}\sigma\bm\zeta\otimes\bm\zeta}.
\end{align}

After calculating the Gaussian integral in Eq.~(\ref{expression-omega-G}) over $(2N-4)$ modes
$\bm\zeta$ with positive eigenvalues at the saddle point (cf. Eq. (\ref{functional-Arhodet})), one derives
%-------------------------------------------------------------------------------------
\begin{eqnarray}
\label{omega-J0-result}
\omega= J_{0}\vartheta_{0}N^N(\pi\kappa)^{N-2}/{\sqrt{\det\sigma}}\,.
\end{eqnarray}
%-------------------------------------------------------------------------------------

We combine Eqs. (\ref{normalize-multi-result}), (\ref{J0-Z}) and (\ref{omega-J0-result}) to get the
topological current in terms of $Z$:
%-------------------------------------------------------------------------------------
\begin{eqnarray}
\label{omega-Z-result} \omega= \vartheta_{0}/(4\pi^{2}\kappa)\sqrt{\det\gamma/\det\sigma}
\sqrt{\kappa k_{0}/(4\pi)}(Z^{-1} e^{-S_+} - Z e^{-S_-})\,.
\end{eqnarray}
%-------------------------------------------------------------------------------------
This equation has a form
%-------------------------------------------------------------------------------------
\begin{eqnarray}
\label{omega-kappapm} \omega = Z^{-1} \kappa_+ - Z \kappa_-,
\end{eqnarray}
where the newly introduced quantities
%-------------------------------------------------------------------------------------
\begin{eqnarray}
\label{kappapm-S12}
\kappa_\pm=\vartheta_{0}/(4\pi^{2}\kappa)\sqrt{\det\gamma/\det\sigma}
\sqrt{\kappa k_{0}/(4\pi)}e^{-S_{\pm}}
\end{eqnarray}
can be interpreted as
the transition rates in the auxiliary Markov chain model discussed in detail in the next Subsection
\ref{subsubsec:markov-chain}. The stationary current can be
obtained by setting $Z=1$. The final expression of $Z$ via $\omega$ follows from Eq. (\ref{omega-kappapm}) (the
choice of the root is related to the model equivalence discussed in Appendix
\ref{app:markov-chain}):
%-------------------------------------------------------------------------------------
\begin{eqnarray}
\label{Z-omega} Z = \frac{\sqrt{\omega^2+4\kappa_+\kappa_-} - \omega}{2\kappa_-}\,.
\end{eqnarray}
%-------------------------------------------------------------------------------------

\subsubsection{Final expressions and two-channel single-state Markov chain}
\label{subsubsec:markov-chain}

We complete the derivation of the Cram\'er function by providing the relation
to the parameters of the spin-chain model $v$ and $u$.
The ratio of the determinants can be calculated in the limit $N\gg 1$ (see
Appendix~\ref{subapp:determinants} for details):
%-------------------------------------------------------------------------------------
\begin{eqnarray}
\label{detdet} \sqrt{\det \gamma/\det \sigma} = \pi^{-2}v^4 u^{-1}(v^4-u^4)^{-1}\sin(\pi u/v)\,.
\end{eqnarray}
%-------------------------------------------------------------------------------------

The final expression for the Cram\'er function is
%-------------------------------------------------------------------------------------
\begin{eqnarray}
\label{functional-final}
{\cal S} = -\omega\ln\frac{\sqrt{\omega^2+4\kappa_+\kappa_-} - \omega}{2\kappa_-}
 -\sqrt{\omega^2+4\kappa_+\kappa_-} + \kappa_+ + \kappa_-\,,
\end{eqnarray}
%-------------------------------------------------------------------------------------
where $\kappa_\pm$ are expressed through $v$ and $u$ with the help of
Eqs. (\ref{k0}), (\ref{int-zero-modes-G}), (\ref{kappapm-S12}), and (\ref{detdet}),
whereas the barriers are
%-------------------------------------------------------------------------------------
\begin{eqnarray}
\label{S1S2}
S_{\pm}=2\pi(v \mp u )^{2}/(\kappa v) \,.
\end{eqnarray}
%-------------------------------------------------------------------------------------

%%%%%%%%%%%%%%%%%%%%%%%%%%%%%%%%%%%%%%%%%%%%%%%%%%%%%%%%%%%%%%%%%%%%
\begin{figure}[htp]
  \begin{center}
      \includegraphics[width=0.10\textwidth]{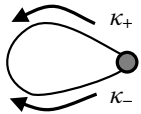}
  \end{center}
  \caption{Two-channel single-state Markov chain.}
  \label{markovsite}
\end{figure}
%%%%%%%%%%%%%%%%%%%%%%%%%%%%%%%%%%%%%%%%%%%%%%%%%%%%%%%%%%%%%%%%%%%%%

Now we can demonstrate the equivalence of the spin-chain model to the simple Markov chain model,
shown in Fig. \ref{markovsite} and described in Appendix \ref{app:markov-chain}, and justify that
$\lambda$ can be neglected in the above solution. Using the properties of the distributions $\bm
J(\bm x)$ and $\bm A(\bm x)$, we can rewrite the first representation of the Cram\'er function in
Eq. (\ref{functional-simplified}) as
%-------------------------------------------------------------------------------------
\begin{eqnarray}
\label{functional-lambdaZ}
{\cal S}= -\omega\ln Z - \lambda
\,.
\end{eqnarray}
%-------------------------------------------------------------------------------------
The eigenvalue $\lambda$ can be calculated by comparing the two representations
of the Cram\'er function:
\begin{align}
\label{lambda-Z}
& \lambda =
Z^{-1}\kappa_+ + Z\kappa_- - (\kappa_+ + \kappa_-)\,,
\end{align}
or in terms of $\omega$,
\begin{align}
\label{lambda-omega}
& \lambda = \sqrt{\omega^2+4\kappa_+\kappa_-} - \kappa_+ - \kappa_-
\,.
\end{align}
The equivalence of the spin-chain model at moderate topological currents
to the two-channel single-state Markov chain model is,
in particular, obvious from the forms of
Eqs. (\ref{Z-omega}), (\ref{functional-lambdaZ}), and (\ref{lambda-Z}).

We see that $\lambda$ can indeed be neglected if the total current is not
exponentially larger than the equilibrium current.

\section{Summary and Conclusions}
\label{sec:conclusions}

Let us briefly summarize main steps, results of the general approach and the model discussed in the manuscript.

We have developed here a topological picture of generated stochastic currents, where each
current component is associated with a topologically nontrivial $1$-cycle in the system
configuration space, so that the current ${\bm\omega}\in H_{1}(M;\mathbb{R})$ resides in the first
homology of the configuration space with real coefficients. The current, defined in a topological
way as the set of number of rotations over the independent cycles per unit time, is related to the
current density ${\bm J}$ via the Poincar\'e duality $H_{1}(M;\mathbb{R})\cong H^{m-1}(M;\mathbb{R})$,
where the cohomology is considered in the de Rahm representation. ( See. e.g., \cite{G-H-alg-geom}
for review of the de Rahm cohomology.) By considering the current density ${\bm J}$ as a
closed form of rank $(m-1)$, the current is viewed ${\bm\omega}=[{\bm J}]$ as its cohomology class.
This allows the Cram\'er function ${\cal S}({\bm\omega})$ of the long-time current distribution to
be calculated by applying the variational principle to the current-density functional ${\cal
S}({\bm J},\rho)$. The explicit form of ${\cal S}({\bm J},\rho)$ together with the variational
principle and an intuitive path-integral base derivation is also presented in the manuscript.

We further focused on the case of topological driving, when the driving force is locally potential,
$d{\bm F}=0$. Even though the system may not be represented by a potential function due
to presence of topologically nontrivial cycles, it still allows description in terms of a multi-valued potential
${\bm F}=-dV_{WZ}$, referred to as a Wess-Zumino potential, due to it close resemblance
with multi-valued Wess-Zumino actions
well-known in quantum field theory. By applying the variational principle to the
calculation of the Cram\'er function ${\cal S}({\bm\omega})$ we derived a system of equations for
$\rho$, ${\bm J}$, and ${\bm A}$, the latter being an auxiliary curvature-free ($d{\bm A}=0$) gauge
field. We also developed a procedure of solving the aforementioned
equations step-by-step, resulting in the well defined algorithm for calculating the
Cram\'er function ${\cal S}({\bm\omega})$ of the topological current.

To illustrate the general approach we considered a circular spin-chain stochastic model with
topological driving that constitutes a regularized version of a $(1+1)$ stochastic field
theory. The model represents Langevin dynamics of an elastic string evolving over a two-dimensional sphere
$S^{2}$. Despite of its high intrinsic dimensionality the problem appears tractable as it has only one
independent topologically nontrivial cycle, $H_{1}(M;\mathbb{R})\cong \mathbb{R}$, which
reflects the well-known result on the homology of the loop spaces of spheres. In the low-noise
limit we have solved aforementioned variational equations for $\rho$, ${\bm J}$, and ${\bm A}$
explicitly. The solvability became possible via proper use of the rare-event,
instanton analysis of the current generation. Moreover,  the actual calculations are streamlined even further,
as in fact we do not perform the path-integral calculations around the instanton solutions explicitly, but rather
use the instanton scenario to build an asymptotically exact ansatz for $(\rho,{\bm J},{\bm A})$.
This useful technical trick allows us to solve the equations (in the proper low noise and large deviation limit)
analytically. In particular, we have found that the current density ${\bm J}$, generating the topological current
$\omega$, is concentrated in a small tubular neighborhood of the finite-dimensional space $M_{0}\subset M$
of instanton configuration, and it shows a Gaussian dependence on the set ${\bm\zeta}$ of the transverse variables.
Technically, this approach can be
viewed as an extension of the equilibrium techniques based on current tubes
\cite{CaroliCaroliRouletGouyet80,LandauerSwanson,CDS08,MaierStein,DMRH,DLMcCS,04TNK}
to the non-equilibrium problem of evaluating
the Cram\'er functions of generated stochastic currents. It is instructive to note that, although
the relevant distributions are concentrated in a small tubular neighborhood of the whole instanton
space $M_{0}$, the calculation of ${\cal S}(\omega)$ requires careful consideration of a small
neighborhood of the saddle point (transition state) only. This reflects the fact that,
during a rare event that contributes to the
current generation, the system spends most of the time around the saddle point, being thrown there
by a strong fluctuation of noise and waiting for a small extra kick that starts its unavoidable
fall-down back to the steady state. Most of the time the spin-chain is involved in a ``boring",
i.e. typical,  diffusive meandering along the sphere, thus waiting (almost forever) for the
next instanton jump/transition. This scenario allows us to reduce the complex stochastic field theory
to a simple Markov chain model.

This asymptotic (low-noise, instanton) reduction of a continuous stochastic system to a simple
Markov chain model in the case of moderate non-equilibrium currents is quite a general result. In
the case of a purely potential force ${\bm F}=-dV$ it is connected to the topological properties of
the potential and closely related to the Morse theory \cite{Hirsch}. A connection between the Morse
theory and instantons was established in \cite{82Wit}, where a super-symmetric imaginary-time quantum
mechanics has been introduced with the effective potential $V_{{\rm eff}}=(dV)^{2}$. In the
semiclassical limit $\hbar\to 0$ the instanton approach has been implemented to evaluate the
effects of tunneling between the metastable states, thus allowing us to distinguish between the true
zero modes, which contribute to the de Rahm cohomology \cite{G-H-alg-geom,Manin} and hence
describe the topological invariants of the configuration space, and just soft modes with
exponentially vanishing eigenvalues in the $\hbar\to 0$ limit. This approach is known as the
Morse-Witten (MW) theory \cite{82Wit}. It is well known \cite{vanKampen}
that a simple gauge transformation turns
the Fokker-Planck (FP) operator into the Schr\"odinger operator in the potential $V_{{\rm eff}}$,
with $\hbar^{2}$ representing the temperature in the FP theory. Within this equivalence the
instantons have the same shape, although in the FP picture they play the role of optimal
fluctuations that minimize the Onsager-Machlup action for rare stochastic transitions between the
metastable states. The eigenvalues of the soft modes, describing tunneling in the MW theory,
attain a true physical meaning in the FP picture, since they represent the slowest relaxation
rates, which are due to rare over-the-barrier transitions. Such an approach has been utilized by Tanase-Nicola and Kurchan \cite{04TNK} who studied a supper-symmetric FP theory that extends the standard FP dynamics in the same way as the super-symmetric quantum mechanics extends standard quantum mechanics.

At this point we note that the standard Morse theory describes the case of a purely potential
force, whereas for our spin-chain model the force $d{\bm F}=0$ is only locally potential. The
topological counterpart of stochastic dynamics in this situation is known as the Morse-Novikov (MN)
theory that has demonstrated its capabilities to study topological properties of the underlying
configuration space \cite{82Nov}. It is worth noting that in the continuous limit the multi-valued
potential $V_{WZ}({\bm n})$ of our spin-chain describes the multi-valued action of a free particle
moving along the sphere $S^{2}$ in a magnetic field of a constant curvature. If the particle motion is affected additionally to
the magnetic field by a non-zero potential, the problem, which can be mapped
onto the Kirchhoff equations, can be handled using the MN theory \cite{82Nov}. In a generic situation
the instanton space $M_{0}$ is one-dimensional and is represented by the unstable spaces of the
isolated critical points of Morse index $1$ (one unstable mode). The spin-chain model, considered
in this manuscript, whose continuous limit corresponds to the Kirchhoff problem with the zero potential,
is degenerate due to its intrinsic $SO(3)$ symmetry, restored when the potential vanishes.
In terms of the Morse theory this degeneracy means that the critical points, which are isolated in
the case of standard Morse (including MN) theory are replaced by isolated critical manifolds associated
with the orbits of the underlying symmetry group, $SO(3)$ in our case. Such a situation, in the
potential force case is described by the so-called equivariant Morse-Bott (MB) theory \cite{82Bot}.
Therefore, the topological counterpart of the low-noise stochastic dynamics for our spin-chain model
can be referred to as a case of the equivariant Morse-Bott-Novikov (MBN) theory. Stated in the stochastic
dynamics terms, the instanton space is represented by a $1$-dimensional family of the $SO(3)$
orbits, actually $SO(3)$ itself, almost everywhere. Then only the stable points (constant loops)
are represented by an orbit from $S^{2}\cong SO(3)/SO(2)$. However, and in
spite of this degeneracy, we have just showed that this collapse of the orbits at the stable
critical points is not an obstruction for the general Cram\'er function calculations. Let us also
emphasize that even though traditionally the standard (equivariant) Morse theory was developed
primarily for the case of a potential force, we found out that in fact it can be efficiently
used in the stochastic dynamics aspect in the case of moderate topological driving, when the Morse
function is multi-valued. As we just demonstrated on the example of the non-equilibrium
spin-chain, moderate character of the driving does not change the topological structure of the
critical and instanton manifolds in comparison with the pure potential case.

In a generic non-equivariant case with moderate topological driving one could expect a full reduction
of the low-noise stochastic dynamics to a Markov chain process on a graph whose nodes and links
represent
the critical points of Morse index $0$ and the unstable manifolds of the critical points of
Morse index one, respectively. This is really the case when the potential function satisfies the
Morse-Smale (MS) condition \cite{Hirsch}, which would be a generic situation. If the MS condition
is not satisfied (which is some kind of degeneracy) the stochastic dynamics of the current
generation can show some additional interesting features, which are yet to be analyzed. In the
case $d{\bm F}\ne 0$ of intrinsically non-potential force, whose topological counterpart is
represented by Morse decompositions in the Conley index theory \cite{78Con}, stochastic dynamics can be
extremely complicated, with the critical spaces, represented by neither isolated points nor even
isolated manifolds but rather some closed sets, possibly of fractal nature. Even in the simplest
non-equivariant case of isolated critical points, the situation is much more complicated
(than in the equivariant case), yet apparently treatable. The difficulty is that transitions between
different isolated states of the effective Markov chain become direction-dependent, since as opposed
to the potential and locally-potential cases (with isolated critical points) the instanton trajectories
that correspond to climbing the barrier and falling down the barrier are different.
The problem of computing the pre-exponential factors in this case also becomes much more involved and less universal.

Let us also stress that in the $v=0$ case the Onsager-Machlup action of the spin-chain
model in the thermodynamic field theoretic limit, $N\to\infty$, reproduces the action of a
$(1+1)$ sigma-model with the topological term. However, in the regularized model (finite $N$)
the relative fluctuations in the $v=0$ case are strong, so that ${\bm n}_{j+1}$ is typically
not close to ${\bm n}_{j}$. The corresponding string on the sphere is not continuous, which can be
viewed as the main reason for complexity of the sigma-model, considered as a field theory.
The elastic term proportional to $v$ suppresses the relative fluctuations
in such a dramatic
way that statistically ${\bm n}_{j+1}$ and ${\bm n}_{j}$ are always close to each other.
Therefore, the configurations can be viewed as continuous loops in $S^{2}$ even
for a finite large $N$, which creates a topologically nontrivial cycle.
This interesting  peculiarity of the problem gives rise to generation of the single component
stochastic current. We also  note that since in the $v\ne 0$ case and $N\to\infty$ the result
is finite, the elastic term in the case
of small $v$ can be interpreted as a regularization of the sigma-model based stochastic
field theory suppressing the short-range fluctuations and eliminating divergences.

The methodology for the LDP of empirical currents, developed in this manuscript, is closely related to the FP approach to MW theory developed by Tanase-Nicola and Kurchan \cite{04TNK}. In a way, this methodology can be interpreted as an extension of the approach of \cite{04TNK} to the locally potential driving case $d{\bm F}=0$, implemented in low-dimensional $0$- and $1$-fermion sectors that correspond to  the density and current density distributions, respectively, and applied to large-time statistics of empirical currents. Thus, we believe that our weak noise analysis of currents in the problems with topological driving can be restated in super-symmetric terms of \cite{04TNK}. Testing this conjecture remains a future challenge. It would also be interesting to extend the above relation to higher dimensions, e.g., by treating diffusion in $n$-fermion sector as noisy dynamics of $n$-dimensional chains that produces higher dimensional currents, residing in $H_{n+1}(M;\mathbb{R})$. Based on the presented in the manuscript low-dimensional results, it appears that the low-noise limit of the Cram\'er functions of the empirical currents (including the higher-dimensional counterparts) contains detailed information on the Morse decomposition of the underlying space $M$.

Finally, we have considered the simplest stochastic $(1+1)$ sigma model with the target space
represented by $S^{2}\cong \mathbb{CP}^{1}$. A simple topological computation demonstrates that a generalization
based on the target spaces $\mathbb{CP}^{n}$ with $n>1$ still provide a single component
topological current. The simplest generalization resulting in the multi-component topological
currents ${\bm\omega}$ requires the target space to be a complex flag space
\cite{G-H-alg-geom,Manin}. The Cram\'er function ${\cal S}({\bm\omega})$ can be computed in
such cases straightforwardly and without any significant complications,
by using the methodology explained above.

\acknowledgments
It is our pleasure to thank Reviewer 2 for an overwhelmingly deep, extensive, and detailed report, as well as a number of interesting and useful comments.
This material is based upon work supported by the National Science Foundation
under Grant No. CHE-0808910. Research at LANL was carried out under the auspices of the National
Nuclear Security Administration of the U.S. Department of Energy at Los Alamos National Laboratory
under Contract No. DE C52-06NA25396.

\appendix

\section{Topological structure of the configuration space}
\label{app:top-config-space}

In this Appendix we discuss mathematical formulations for the spin chain regularization introduced in
Section \ref{subsec:SpinChainTop} in a somewhat lighter, more physical terms.

Let us first introduce a simple and useful  representation of the first homotopy and homology
groups of the relevant configuration space ${\rm Map}(S^{1},S^{2})$  of maps from the base space
$S^{1}$ to $S^2$. To regularize our statistical field theory we should consider the spaces of maps
$S^{1}\to S^{2}$, with various degrees of smoothness. Ultimately, we are interested in random
maps (non-smooth) whose topological properties can be approximated using
well-behaved (continuous) maps.

The spaces of smooth and piece-wise smooth maps from $S^1$ to $S^2$ are denoted by ${\rm
Map}^{\infty}(S^1,S^2)$ and ${\rm Map}_{{\rm p}}^{\infty}(S^1,S^2)$, respectively.  The topology of
these map spaces is defined in a standard way. To analyze proper discretizations of the spaces of
smooth maps, we represent the circle $S^1$ by a cyclic lattice $j=0,\ldots,N-1$, approximate a map
${\bm n}(y)$ by a set of points ${\bm n}_{j}$, and introduce $\varepsilon=2\pi/N$. Finally, we
define a set of approximations $L_{N,\varepsilon_{0}}S^2$ for the space ${\rm Map}(S^1,S^2)$ by
%-------------------------------------------------------------------------------------
\begin{eqnarray}
\label{define-approx-space} L_{N,\varepsilon_{0}}S^2=\left\{{\bm n}\in (S^2)^{\times N}|\;1-{\bm
n}_{j+1}\cdot{\bm n}_{j}<\varepsilon_{0}, \; \forall j=0,\ldots,N-1 \right\}.
\end{eqnarray}
%-------------------------------------------------------------------------------------
The elements of $L_{N,\varepsilon_{0}}S^2$ are represented by $N$-tuples of points in $S^2$ such
that the neighboring points are sufficiently close to each other. For not too large
$\varepsilon_{0}$, e.g., for $\varepsilon_{0}<1/3$, we can define a continuous map
$L_{N,\varepsilon_{0}}S^2\to {\rm Map}(S^1,S^2)$ by connecting the neighboring points ${\bm
n}_{j}$, ${\bm n}_{j+1}$ with geodesic lines.

This map generates homomorphisms between the homotopy, homology, and cohomology groups,
respectively, for all the three spaces of maps.  It is possible to show that if an approximation
is accurate enough, the relevant topological properties of the approximations
$L_{N,\varepsilon_{0}}S^2$ are identical to these of the original space ${\rm Map}(S^1,S^2)$.

To identify a very simple and intuitive picture of the topological current generation in our field
theory, we define continuous maps $\theta:M\to S^1$ for $M={\rm Map}^{\infty}(S^1,S^2)$, $M={\rm
Map}_{{\rm p}}^{\infty}(S^1,S^2)$, and $M=L_{N,\varepsilon_{0}}S^2$ by associating with any loop
from $M$ its {\emph {Berry phase}} defined as the holonomy along ${\bm x}\in M$, also understood
as ${\bm x}:S^{1}\to S^{2}$.

Introducing a map from $e^{i\varphi_B/2}\in U(1)$ to $\theta \in [0,2\pi]\to
S^1$ and combining it with the maps defined above, we arrive at the continuous maps
$\theta:M\to S^1$ for $M={\rm Map}^{\infty}(S^1,S^2)$,
$M={\rm Map}_{{\rm p}}^{\infty}(S^1,S^2)$, and $M=L_{N,\varepsilon_{0}}S^2$. In all three cases the
morphisms $\pi_{1}(M)\to \pi_{1}(S^1)\cong {\mathbb Z}$ and $H_{1}(M)\to H_{1}(S^1)\cong {\mathbb
Z}$ generated in the homotopy $\pi_{1}$ and homology $H_{1}$ groups, respectively, are
isomorphisms. All this implies that the current generation can be observed by monitoring
the reduced variable $\theta({\bm x})\in S^1$ and counting the windings around the circle.

\section{Derivation of the Cram\'er function for the spin-chain model: details}
\label{app:spin-chain-Cramer-detail}

In this Appendix we present some details of the Cram\'er function ${\cal
S}({\bm\omega})$ derivation for the spin-chain model.

\subsection{Vector potential in the WKB region}
\label{subapp:WKB-A}

In the WKB region we set $\lambda=0$ and neglect the nonlinear term in Eq.~(\ref{S-omega-eq-1}),
which turns it into
%-------------------------------------------------------------------------------------
\begin{eqnarray}
\label{eq-A-WKB} (\kappa/2)d^{\dagger}{\bm A}+{\bm F}\cdot{\bm A}=0, \;\;\; d{\bm A}=0.
\end{eqnarray}
%-------------------------------------------------------------------------------------
Since the WKB region does not contain topologically nontrivial $1$-cycles,
Eq.~(\ref{eq-A-WKB}) can be recast in the following form:
%-------------------------------------------------------------------------------------
\begin{eqnarray}
\label{eq-A-WKB-1} {\cal L}^{\dagger}\psi_{-}=0, \;\;\; {\bm A}=d\psi_{-}={\bm\partial}\psi_{-}.
\end{eqnarray}
%-------------------------------------------------------------------------------------
We seek the solution of this equation in the following form
%-------------------------------------------------------------------------------------
\begin{eqnarray}
\label{eq-A-WKB-2} \psi_{-}({\bm x})=\psi({\bm x})e^{2\kappa^{-1}V_{WZ}({\bm x})}, \;\;\; \psi({\bm
x})=\psi_{0}(\theta)e^{-\kappa^{-1}\sigma(\theta)({\bm\zeta}\otimes{\bm\zeta})}, \;\;\; {\cal
L}\psi({\bm x})=0,
\end{eqnarray}
%-------------------------------------------------------------------------------------
or, equivalently,
%-------------------------------------------------------------------------------------
\begin{eqnarray}
\label{eq-A-WKB-3} \psi_{-}({\bm
x})=\psi_{0}(\theta)e^{2\kappa^{-1}V_{0}(\theta)-\kappa^{-1}\sigma_{-}(\theta)({\bm\zeta}\otimes{\bm\zeta})},
\;\;\; \sigma_{-}(\theta)=\sigma(\theta)-W(\theta).
\end{eqnarray}
%-------------------------------------------------------------------------------------
We further substitute $\psi({\bm x})$ given by Eq.~(\ref{eq-A-WKB-2}) in the equation ${\cal
L}\psi({\bm x})=0$, retain only the leading terms in $\kappa$, and keep in mind that typically
$|{\bm\zeta}|\sim\sqrt{\kappa}$. Overall this results in the following system of equations:
\begin{eqnarray}
\label{sigma-psi-minus-eq}  -\partial_{\theta}(F_{0}\psi_{0})-\psi_{0}{\rm Tr}(\sigma-W)=0, \;\;\;
-F_0\nabla_{\theta}\sigma=2\sigma^{2}-W\sigma-\sigma W.
\end{eqnarray}
%-------------------------------------------------------------------------------------
Combining the two  Eqs.~(\ref{sigma-psi-minus-eq}) one derives
%-------------------------------------------------------------------------------------
\begin{eqnarray}
\label{sigma-psi-minus-eq-2}
\partial_{\theta}( F_{0}\psi_{0}/\sqrt{\det\sigma})=0, \;\;\;
-F_0\nabla_{\theta}\sigma=2\sigma^{2}-W\sigma-\sigma W.
\end{eqnarray}
%-------------------------------------------------------------------------------------

The second equation in Eqs.~(\ref{sigma-psi-minus-eq-2}) is nonlinear, and thus intractable, if the matrix is large.
However, one can still show that this equation does have a solution with the following important property,
$\sigma(\theta)=W(\theta)$ for $\theta=0,\,
\theta_{0},\, 2\pi$. To demonstrate this property one refers to a dynamical equations with respect
to some proper time $\tau$
%-------------------------------------------------------------------------------------
\begin{eqnarray}
\label{sigma-eq-dynamical} \dot{\theta}=F_{0}(\theta), \;\;\; \dot{\sigma}=-2\sigma^{2}+\sigma
W(\theta)+W(\theta)\sigma, \;\;\; \dot{\theta}\equiv \partial_{\tau}\theta, \;\;\;
\dot{\sigma}\equiv\nabla_{\tau}\sigma.
\end{eqnarray}
%-------------------------------------------------------------------------------------
This dynamical system has three critical points
$(\theta,W(\theta))$ with $\theta=0,\, \theta_{0},\, 2\pi$, where $\theta=0,\, 2\pi$ correspond to
stable critical points, while $\theta=\theta_{0}$ describes an unstable critical point
($\partial_\theta F_0(\theta_0) > 0$).
Therefore, finding a solution with the desired properties is equivalent to identifying  two solutions of
Eq.~(\ref{sigma-eq-dynamical}) which start at the unstable point and reach the two stable points
in infinite time. Generally such trajectories do exist. For our spin-chain model it can be
demonstrated in a straightforward manner, e.g., by using the angular-momentum representation for
the transverse modes $\psi_{j}^{a}(y;\theta)$ entering Eq.~(\ref{coord-U-explicit}).

The first equation in Eqs.~(\ref{sigma-psi-minus-eq-2}) can be easily solved:
%-------------------------------------------------------------------------------------
\begin{eqnarray}
\label{psi-0-WKB} \psi_{0}(\theta)=C\sqrt{\det\sigma(\theta)}/F_{0}(\theta),
\end{eqnarray}
%-------------------------------------------------------------------------------------
thus immediately providing expressions for $A_{\theta}$, via Eqs.~(\ref{eq-A-WKB-1}) and
(\ref{eq-A-WKB-3}). Therefore, the WKB solution for $A_{\theta}$ that matches with the harmonic
solution (\ref{A-saddle-solution}) within the domains where both apply, has the following form in
the two distinct sub-domains:
%-------------------------------------------------------------------------------------
\begin{eqnarray}
\label{A-WKB-solution}
&& 0 \lesssim\theta\lesssim \theta_0:\quad A_\theta = -\sqrt{\kappa
k_{0}\det\sigma/(\pi\det W(\theta_{0}))}(1/Z-1)e^{2\kappa^{-1}\left(V_{0}(\theta)-V_{0}(\theta_{0})
\right)}e^{-\kappa^{-1}\sigma_{-}(\theta)({\bm\zeta}\otimes{\bm\zeta})},
\\
&& \theta_{0} \lesssim\theta\lesssim 2\pi:\quad A_\theta = -\sqrt{\kappa
k_{0}\det\sigma/(\pi\det W(\theta_{0}))} (1-Z) e^{2\kappa^{-1}\left(V_{0}(\theta)-V_{0}(\theta_{0})
\right)}e^{-\kappa^{-1}\sigma_{-}(\theta)({\bm\zeta}\otimes{\bm\zeta})}.
\end{eqnarray}
%-------------------------------------------------------------------------------------
The expressions for the transverse components $A_{j}$ of the vector potential can also be derived straightforwardly.

\subsection{Current tubes in the WKB region}
\label{subapp:WKB-J}

In this Subsection we determine the current density distribution ${\bm J}$. To achieve this goal we
solve Eq.~(\ref{S-omega-eq-2}) assuming that ${\bm J}$ is concentrated in a small neighborhood of
the instanton space $M_{0}$.

In the WKB region we can set ${\bm A}=0$ in the first equation in Eqs.~(\ref{S-omega-eq-2}), and
since the region does not contain topologically nontrivial $1$-cycles, the first equation is
equivalent to ${\bm J}=(\kappa/2)d\varphi-{\bm F}\varphi$ for some function $\varphi({\bm x})$.
Substituting this representation into the second equation of Eqs.~(\ref{S-omega-eq-2}) one arrives at the
following system of equations for ${\bm J}$
%-------------------------------------------------------------------------------------
\begin{eqnarray}
\label{S-omega-eq-2-a} {\cal L}\varphi({\bm x})=0, \;\;\; {\bm
J}=(\kappa/2){\bm\partial}\varphi-{\bm F}\varphi.
\end{eqnarray}
%-------------------------------------------------------------------------------------
In the low-noise limit the current is generated by rare events,
and during the transition the system experiences small
Gaussian fluctuations around the instanton trajectories. Therefore, the current density ${\bm J}$
is concentrated in a small tubular neighborhood $U_{{\rm curr}}\supset M_{0}$ of the instanton
manifold $M_{0}$ where its longitudinal current component should have a Gaussian dependence on the
transverse variables ${\bm\zeta}$. To verify this property we seek the solution of
Eq.~(\ref{S-omega-eq-2-a}) in a form
%-------------------------------------------------------------------------------------
\begin{eqnarray}
\label{varphi-ansatz-WKB} \varphi({\bm x})=\varphi(\theta,{\bm\zeta})
=\varphi_{0}(\theta)e^{-\kappa^{-1}\sigma(\theta)({\bm\zeta}\otimes{\bm\zeta})}.
\end{eqnarray}
%-------------------------------------------------------------------------------------
We substitute the ansatz of Eq.~(\ref{varphi-ansatz-WKB}) into Eq.~(\ref{S-omega-eq-2-a}) and apply
the same strategy as used earlier in Section~\ref{subapp:WKB-A} in the context of deriving
Eqs.~(\ref{sigma-psi-minus-eq}) and (\ref{sigma-psi-minus-eq-2}). This results in the following
system of equations
%-------------------------------------------------------------------------------------
\begin{eqnarray}
\label{sigma-varphi-eq}
\partial_{\theta}( F_{0}\varphi_{0}/\sqrt{\det\sigma})=0, \;\;\;
-F_0\nabla_{\theta}\sigma=2\sigma^{2}-W\sigma-\sigma W.
\end{eqnarray}
%-------------------------------------------------------------------------------------
Since the second relation in Eqs.~(\ref{sigma-psi-minus-eq-2}) and (\ref{sigma-varphi-eq}) are
identical, using the same notation $\sigma$ for the variances in Eqs. (\ref{eq-A-WKB-2}) and
(\ref{varphi-ansatz-WKB}) is perfectly legitimate.

Solving the first equation in Eq.~(\ref{sigma-varphi-eq}), further substituting the obvious
solution into Eq.~(\ref{S-omega-eq-2-a}), and keeping the leading terms in $\sqrt{\kappa}$ in the
way detailed in Section~\ref{subapp:WKB-A}, we arrive at the following expression for the
current density ${\bm J}$ in the WKB region:
%-------------------------------------------------------------------------------------
\begin{eqnarray}
\label{J-expression-WKB} J_{\theta}(\theta,{\bm\zeta})&=&J_{0}\sqrt{\det\sigma(\theta)/\det
W(\theta_{0})}e^{-\kappa^{-1}\sigma(\theta)({\bm\zeta}\otimes{\bm\zeta})}, \nonumber \\
J_{i}(\theta,{\bm\zeta})&=&J_{0}\sqrt{\det\sigma(\theta)/\det
W(\theta_{0})}(F_{0}(\theta))^{-1}(\sigma_{ij}(\theta)-W_{ij}(\theta))\zeta_{j}
e^{-\kappa^{-1}\sigma(\theta)({\bm\zeta}\otimes{\bm\zeta})}.
\end{eqnarray}
%-------------------------------------------------------------------------------------
Careful examination of the second equation in Eqs.~(\ref{sigma-varphi-eq}) in the harmonic region
of the saddle point and allowing for the properties of the solution described at the end of
Section~\ref{subapp:WKB-A} show that $\sigma_{-}(\theta)=\sigma(\theta)-W(\theta)$ tends to
zero, when $\theta$ approaches $\theta_{0}$, as $\sim(\theta-\theta_{0})^{2}$. Therefore, the
transverse components $J_{j}$ vanish in the harmonic region. This confirms the assumption we have
made in Section~(\ref{subsec:derive-Cramer}), referred to there as the property (d).

At this point we note that the current distribution ${\bm J}$, given by
Eq.~(\ref{J-expression-WKB}) derived for the WKB region, also extends nicely into the harmonic region
of the saddle point $\theta=\theta_{0}$, thus suggesting that ${\bm J}$ in
this region is also described by Eq.~(\ref{J-expression-WKB}). This can be verified directly. The
reason why the WKB solution easily extrapolates into the harmonic region is related to the asymptotically
longitudinal nature of the vector potential ($A_{j}=0$) in the harmonic region of the saddle point.

\subsection{Computation of the relevant determinants}
\label{subapp:determinants}

In this Section we present some details of the functional integral calculation in Eqs.
(\ref{normalize-rho}), (\ref{functional-Arhodet}), and (\ref{omega-J0}). In Eq.
(\ref{normalize-rho}) the integration is performed over the deviations from the stable constant
loop, whereas both in Eq. (\ref{functional-Arhodet}) and Eq. (\ref{omega-J0}) the integral runs over
the transverse deviations from the saddle-point loop. In the following calculation, as in the main
text, both the density $\rho$ and the current density $\bm J$ are assumed regularized on a
lattice of $N$ spins, in accordance with discussion of Appendix \ref{app:top-config-space}. Therefore, the
functional integrals should be represented by finite-dimensional integrals over $N$ positions ${\bm
n}_{j}$ on the unit sphere. Since we will see that only long-wavelength deviations from the constant and
saddle-point loops contribute to the ratio of the two integrals of interest,
we will simply execute the limit $N\to\infty$ in all the intermediate expressions
where it exists.

The integral in Eq. (\ref{normalize-rho}) is evaluated over the two-parametric deviations
$\delta n_\alpha(y)$ (with $\alpha=1,2$) from the constant loop.
The potential accounting for configurations around the constant loop is harmonic:
%-------------------------------------------------------------------------------------
\begin{eqnarray}
\label{potential-at-stable} V_{WZ} \approx
(1/2) (\delta\bm n, \gamma\delta\bm n)
\,,
\qquad
\gamma=2\pi\left(
         \begin{array}{cc}
           -v\partial_y^2 & u\partial_y \\
           -u\partial_y & -v\partial_y^2 \\
         \end{array}
       \right)
\,,
\end{eqnarray}
%-------------------------------------------------------------------------------------
where the scalar product is conventionally defined as
%-------------------------------------------------------------------------------------
\begin{eqnarray}
\label{scalar-product}
(\bm\xi, \bm\eta)=\int_0^{2\pi}\frac{dy}{2\pi}\sum_{\alpha}\xi^*_\alpha(y) \eta_\alpha(y) \,.
\end{eqnarray}
%-------------------------------------------------------------------------------------

The operator $\gamma$ can be diagonalized in the space of Fourier harmonics $e^{iqy}$ with $q=0,\pm
1,\pm 2,\ldots,\pm (N-1)/2$. For the sake of convenience, and since it does affect the $N\to\infty$ limit,
we consider an odd $N$. The eigenvalues and the corresponding normalized eigenvectors of $\gamma$ are
%-------------------------------------------------------------------------------------
\begin{align}
\label{gamma-eigenvalues}
&
\gamma_q=2\pi(vq^2+uq) \ \ \mathrm{ and } \ \  \bar\gamma_q=2\pi(vq^2-uq)\,,
\\
\label{gamma-eigenvectors}
& \bm E_q = (1/\sqrt{2})e^{iqy}\left(
         \begin{array}{c}
           1 \\
           -i \\
         \end{array}
       \right)
 \ \ \mathrm{ and } \ \
\bar{\bm E}_q = (1/\sqrt{2})e^{iqy}\left(
         \begin{array}{c}
           1 \\
           i \\
         \end{array}
       \right)
\,.
\end{align}
There are two zero eigenmodes in this set: $\bm E_0$ and $\bar{\bm E}_0$. All other eigenvalues are
positive in the considered case $v>u$.
%-------------------------------------------------------------------------------------
One arrives at the following expansion in terms of the eigenvectors:
%-------------------------------------------------------------------------------------
\begin{eqnarray}
\label{F-series-minimum}
\delta\bm n=(1/\sqrt{2N})\sum\limits_q (c_q\bm E_q  +\bar{c}_q \bar{\bm E}_q)\,.
%\delta\bm n=(1/\sqrt{N})\sum\limits_q (c_q\bm E_q  +\bar{c}_q \bar{\bm E}_q)\,.
\end{eqnarray}
%-------------------------------------------------------------------------------------
Since $\delta\bm n$ is real and $\bm E^*_q = \bar{\bm E}_{-q}$,
the coefficients are related as $\bar{c}_{-q}=c_q^*$, and the transformation from the set of $2N$
spin vector components $\{(\delta n_1(y),\delta n_2(y))\}$ with $y=0,\varepsilon, \ldots, 2\pi-\varepsilon$
(where $\varepsilon = 2\pi/N$) to
the set of $N$ Fourier harmonic components $\{(\mathrm{Re} c_q, \mathrm{Im} c_q)\}$ with $q=0,\pm 1, \ldots$
has the Jacobian equal to unity.
The zero mode coordinates $\{\mathrm{Re} c_0, \mathrm{Im} c_0\}$
are related to a uniform shift $\delta \bm n = {\rm const}$ along the sphere as
$c_0 = \sqrt{N} (\delta n_1 + i\delta n_2)$.
%-------------------------------------------------------------------------------------
Thus, the integral over all deviations $\delta\bm n$ from the constant loop consists of a zero-mode factor
$4\pi N$
and $2(N-1)$ integrals over positive modes $\bm\xi$, finally giving
%-------------------------------------------------------------------------------------
\begin{align}
\label{integral-xi} & \int_{S^2}d\mu({\bm n})\int{\cal D}{\bm\xi} \exp\left(
-\kappa^{-1}(\bm\xi, \gamma\bm\xi)\right)= 4\pi^{N}N^N\kappa^{N-1} /\sqrt{\det\gamma}
\,,
\end{align}
%-------------------------------------------------------------------------------------
where the determinant $\det \gamma$ includes $2(N-1)$ positive eigenvalues of $\gamma$.

The integral in Eqs. (\ref{functional-Arhodet}) and (\ref{omega-J0})
over transverse deviations $\bm\zeta$ from the saddle-point loop $\bm n_0(y)$ and its SO(3) rotations
can be calculated in a similar fashion. All small deviations,
including the longitudinal ones and rotations, can be decomposed into meridional and zonal components:
%-------------------------------------------------------------------------------------
\begin{eqnarray}
\label{define-deltan-saddle}
&&\delta{\bm n}(y)=\delta n_{1}(y)\hat\theta(y)
+\delta n_{2}(y)\hat\phi(y)\,,
\\
\label{define-meridional-zonal}
&&
\hat\theta(y)={\bm e}_1\cos(\theta_0/2)\cos y - {\bm e}_2\cos(\theta_0/2)\sin y - {\bm e}_3\sin(\theta_0/2)\,,
\quad
\hat\phi(y)={\bm e}_1\sin y + {\bm e}_2\cos y,
\end{eqnarray}
%-------------------------------------------------------------------------------------
with the unit vectors ${\bm e}_i$ introduced before Eq. (\ref{coord-U-explicit}).
The potential expansion of Eq. (\ref{expand-V-WZ}) can be represented in terms of the scalar product
defined in Eq. (\ref{scalar-product}) as
%-------------------------------------------------------------------------------------
\begin{eqnarray}
\label{potential-at-stable1}
V_{WZ}(\bm n_0 + \delta\bm n)-V_0(\theta_0)= (1/2)(\delta\bm n, \sigma\delta\bm n)\,,
\quad
\sigma=2\pi\left(
         \begin{array}{cc}
           -v\partial_y^2-v\left(1-u^2/v^2\right) & -u\partial_y \\
           u\partial_y & -v\partial_y^2 \\
         \end{array}
       \right)
\,.
\end{eqnarray}
%-------------------------------------------------------------------------------------
Note that in this Appendix $\sigma$ acts in the space of all deviations
including the negative mode $\delta n_1(y) = {\rm const } = \delta\theta/2$,
whereas in the main text
$\sigma$ is restricted to the subspace of transverse deviations corresponding to positive
modes. The eigenvalues and the corresponding normalized eigenvectors of $\sigma$ can be found
similarly to those of $\gamma$:
%-------------------------------------------------------------------------------------
\begin{align}
\label{sigma-eigenvalues}
& \sigma_{q,\pm}=2\pi\left(vq^2-(v/2)\left(1-u^2/v^2\right)
\pm\sqrt{q^2u^2+(v^2/4)\left(1-u^2/v^2\right)^2}\right)\,,
\\
\label{sigma-eigenvectors}
&
\bm e_{q,+} = \frac{e^{iqy}}{\sqrt{\sigma_{q,+}-\sigma_{q,-}}}
\left(
         \begin{array}{c}
           \sqrt{\sigma_{q,+}-vq^2} \\
           iqu/\sqrt{\sigma_{q,+}-vq^2} \\
         \end{array}
       \right)
 \ \ \mathrm{ and } \ \
\bm e_{q,-} = \frac{e^{iqy}}{\sqrt{\sigma_{q,+}-\sigma_{q,-}}}
\left(
         \begin{array}{c}
           \sqrt{vq^2-\sigma_{q,-}}\\
           -iqu/\sqrt{vq^2-\sigma_{q,-}} \\
         \end{array}
       \right)
\,.
\end{align}
%-------------------------------------------------------------------------------------
Deviations from the saddle-point loop are expressed in terms of the eigenvectors as
%-------------------------------------------------------------------------------------
\begin{eqnarray}
\label{F-series-saddle}
\delta\bm n=(1/\sqrt{N})\sum\limits_q (c_{q,+}\bm e_{q,+}  +c_{q,-} \bm e_{q,-})\,.
\end{eqnarray}
%-------------------------------------------------------------------------------------
Since $\bm e^*_{q,+}=\bm e_{-q,+}$ and $\bm e^*_{q,-}=\bm e_{-q,-}$,
one finds $c^*_{q,+}=c_{-q,+}$ and $c^*_{q,-}=c_{-q,-}$.
We can choose $c_{0,\pm}$, as well as $\mathrm{Re}c_{q,\pm}$ and $\mathrm{Im}c_{q,\pm}$
with $q=1,\ldots, (N-1)/2$ (for odd $N$), as $2N$ real normal coordinates.
The Jacobian of the transformation to the normal coordinates is $2^{N-1}$.

The operator $\sigma$ has a negative mode
($\sigma_{0,-} = -4k_0\equiv -2\pi v(1-u^2/v^2)$) related to the longitudinal deviation as
$c_{0,-}=\delta\theta\sqrt{N}/2$.
Three zero modes
\begin{align}
\label{sigma-zero-modes}
\ \ \bm e_{\pm 1,-} = (1/\sqrt{v^2+u^2})e^{\pm iy}\left(
         \begin{array}{c}
           v \\
           \mp iu \\
         \end{array}
       \right)
\quad
\mathrm{and}
\quad
\bm e_{0,+} = \left(
         \begin{array}{c}
           0 \\
           1 \\
         \end{array}
       \right)
\,
\end{align}
correspond to solid rotations of the saddle-point loop around the vectors ${\bm e}_i$  by angles
$\phi_i$, respectively ($i=1,2,3$), which can be identified with the normal coordinates as
\begin{align}
\label{rotations-thru-normcoord}
\phi_1=-\left(2N^{-1/2}v/\sqrt{v^2+u^2}\right)\mathrm{Im} c_{1,-}\,, \quad
\phi_2=\left(2N^{-1/2}v/\sqrt{v^2+u^2}\right)\mathrm{Re} c_{1,-}\,,  \quad
\phi_3= c_{0,+}/(\sqrt{N}\sin(\theta_{0}/2))\,.
\end{align}
Finally, after excluding the integration over the
negative mode by introducing the $\delta$-function $\delta(2c_{0,-}/\sqrt{N})$
in the full normal-mode integral, we obtain the integral over the transverse
deviations from the saddle-point configuration in the following form:
%-------------------------------------------------------------------------------------
\begin{align}
\label{integral-zeta}
& \int_{SO(3)}d\mu_{\theta_0}(g) \int {\cal D}\bm\zeta
\exp\left( -\kappa^{-1}\left(\bm\zeta, \sigma\bm\zeta\right)\right)=
2\pi^N N^N \kappa^{N-2}v^{-3}(v^2+u^2)\sqrt{v^2-u^2}/\sqrt{\det \sigma}
\,,
\end{align}
%-------------------------------------------------------------------------------------
where the determinant $\det \sigma$ includes $2(N-2)$ positive eigenvalues of $\sigma$, or
equivalently, all eigenvalues of $W(\theta_0)$.

Now we are in a position to calculate the ratio of the integrals:
%-------------------------------------------------------------------------------------
\begin{eqnarray}
\label{ratio-of-integrals}
\frac{\int_{SO(3)}d\mu_{\theta_0}(g) \int {\cal D}\bm\zeta
\exp\left( -\kappa^{-1}\left(\bm\zeta, \sigma\bm\zeta\right)\right)}
{\int_{S^2}d\mu({\bm n})\int{\cal D}{\bm\xi} \exp\left( -\kappa^{-1}(\bm\xi, \gamma\bm\xi)\right)}
=
\vartheta_{0}\sqrt{\det\gamma/\det\sigma}/(4\pi^{2}\kappa)
\end{eqnarray}
%-------------------------------------------------------------------------------------
with
%-------------------------------------------------------------------------------------
\begin{align}
\label{app:int-zero-modes-G}
\vartheta_{0}=\int_{SO(3)}d\mu_{\theta_{0}}(g)
= 2\pi^{2}(1+u^{2}/v^{2})\sqrt{1-u^{2}/v^{2}}\,,
\end{align}
%-------------------------------------------------------------------------------------
which, in contrast to the individual integrals and determinants, is $N$-independent in the $N\to\infty$
limit. Indeed, we note that the knowledge of the lowest eigenvalues with $q\ll N$, specified in
Eqs. (\ref{gamma-eigenvalues}) and (\ref{sigma-eigenvalues}), allows us to calculate the ratio of
the determinants by re-grouping the factors
\begin{eqnarray}
\label{ratio-of-determinants}
\sqrt{\det\sigma/\det\gamma} =
\left(1/\sigma_{1,+}\right)\prod_{q=2}^{(N-1)/2}\gamma_{q}\bar\gamma_{q}/\left(\sigma_{q,+}\sigma_{q,-}\right)\,.
\end{eqnarray}
The transformation is legitimate because $\gamma_{q}\bar\gamma_{q}/\left(\sigma_{q,+}\sigma_{q,-}\right)$
approaches unity sufficiently fast as
$q$ increases
if $q\ll N$ and $N\to\infty$.
Then the upper limit of the product can be
extended to infinity. In the meantime, the Weierstrass factorization theorem \cite{TFKP} for the
representation of an entire function as an infinite product leads to the following identity
\begin{eqnarray}
\label{infinite-product}
\prod_{q=2}^{\infty}(1-z^2/q^2) = (\pi z)^{-1}(1-z^2)^{-1} \sin \pi z\,.
\end{eqnarray}
Therefore, applying this formula with $z=u/v$ and $z\to 1$ to Eq. (\ref{ratio-of-determinants}), one finds
\begin{eqnarray}
\label{ratio-of-determinants-result}
\sqrt{\det\sigma/\det\gamma} =
(2\pi)^{-1}v\left(v^2+u^2\right)^{-1}\prod_{q=2}^{\infty}(1-u^2v^{-2}q^{-2})/(1-q^{-2})=
\pi^{-2}v^4 u^{-1}(v^4-u^4)^{-1}\sin(\pi u/v)\,.
\end{eqnarray}

\section{Two-channel single-state Markov chain}
\label{app:markov-chain}

In this Appendix we re-derive the statistics of the current working with the simple stochastic model
consisting only of one site linked to itself by an edge, with jump rates $\kappa_+$ and $\kappa_-$ in the
positive and negative directions, respectively. This simple model, represented in
Fig.~\ref{markovsite}, describes a continuous time random walk \cite{vanKampen}, and reflects the
instanton mechanism of the current generation of our spin-chain model in the weak-noise limit.

A trajectory in the model is specified by a sequence of the jump directions $w_j=\pm 1$ and the
times when they occur: $\bm \eta = \{w_1, \tau_1; w_2, \tau_2; \ldots; w_n, \tau_n \} $, where
$0\leq \tau_1 \leq \tau_2 \leq \ldots \leq t$.

To study the statistics of the jump rate one introduces counting stochastic process
\cite{HSch07} $c(\bm\eta)$ that measures the difference of jump counts in the positive and negative
directions (which is similar to the definition in Eq.~(\ref{rho-jay2})). Its value increases by
$w_{j}=\pm 1$ at the jump $j$ if a jump occurs in the positive or negative direction,
respectively:
%-------------------------------------------------------------------------------------
\begin{align}
\label{randwalk-counting}
&
c(\bm\eta) = \sum_j w_j
\,.
\end{align}
%------------------------------------------------------------------------------------

We are interested in the statistics of the
%time-averaged
empirical current $\omega$, i.e., the number of
jumps per unit time $c(\bm\eta)/t$. The probability of having $\omega t$ jumps during time $t$ is
given by the discrete distribution
%-------------------------------------------------------------------------------------
\begin{align}
\label{randwalk-currentdist}
&
P(\omega,t) = \langle \delta_{\omega t, c(\bm\eta)} \rangle =
\oint_{|Z|=1} \frac{dZ}{2\pi i Z} e^{\omega t \ln Z} \langle e^{-c(\bm\eta)\ln Z} \rangle
\,,
\end{align}
%------------------------------------------------------------------------------------
where $\delta$ is the Kronecker symbol, and the angular brackets denote the average over stochastic
trajectories. This average is determined by the Markovian measure ${\cal
D}{\bm\eta}\exp\left(-S({\bm\eta})\right)$, where ${\cal D}{\bm\eta}=\sum_{n}d\tau_{1}\ldots
d\tau_{n}$ and the action reads
%-------------------------------------------------------------------------------------
\begin{align}
\label{randwalk-pathprob} & \exp\left(-S({\bm\eta})\right)=
\prod_{j=1}^{n}\kappa_{w_j}e^{-(\kappa_+ + \kappa_-)t}\,,
\end{align}
%-------------------------------------------------------------------------------------
with $\kappa_{\pm 1}=\kappa_{\pm}$ being the jump rates.

The generating function $\langle e^{-c(\bm\eta)\ln Z} \rangle$ of the distribution $P(\omega,t)$
can be calculated using the standard procedure, e.g. described in ~\cite{GKP06}:
%-------------------------------------------------------------------------------------
\begin{align}
\label{randwalk-generating}
&
\langle e^{-c(\bm\eta)\ln Z} \rangle = e^{\lambda(Z)t}
\,,
\end{align}
%------------------------------------------------------------------------------------
where
%-------------------------------------------------------------------------------------
\begin{align}
\label{randwalk-lambda}
&
\lambda(Z) = Z^{-1}\kappa_+ + Z\kappa_- - \kappa_+ - \kappa_-
\end{align}
%------------------------------------------------------------------------------------
is the only eigenvalue of the biased $1\times 1$ transition ``matrix'', obtained by replacing the
jump rates as $\kappa_+ \to Z^{-1}\kappa_+$ and $\kappa_- \to Z\kappa_-$, while keeping the overall
escape rate equal to $(\kappa_+ + \kappa_-)$.

Although the distribution can be obtained exactly in terms of the modified Bessel function,
%-------------------------------------------------------------------------------------
\begin{align}
\label{randwalk-exactresult} & P(\omega,t) = I_{\omega t}(2\sqrt{\kappa_+\kappa_-} t) \left(
\kappa_{+}/\kappa_{-}\right)^{\omega t/2}e^{-(\kappa_+ + \kappa_-)t}\,,
\end{align}
%------------------------------------------------------------------------------------
to derive the Cram\'er function ${\cal S}(\omega) = \lim_{t\to\infty}t^{-1} \ln
P(\omega,t)$ we only need to calculate the integral in Eq. (\ref{randwalk-currentdist}) within the
saddle-point approximation. Thus, one derives
%-------------------------------------------------------------------------------------
\begin{align}
\label{randwalk-cramer} & {\cal S}(\omega) = -\lambda(Z) - \omega\ln Z \,,
\end{align}
%------------------------------------------------------------------------------------
where $Z$ is expressed in terms of $\omega$ by means of the saddle-point equation
%-------------------------------------------------------------------------------------
\begin{align}
\label{randwalk-Zomega} & \partial_{Z}{\cal S} \equiv-\partial_{Z}\lambda(Z)-\omega Z^{-1} = 0
\end{align}
%------------------------------------------------------------------------------------
with $\lambda(Z)$ specified in Eq. (\ref{randwalk-lambda}). The resulting quadratic equation for
$Z$ has two real solutions of opposite signs. The positive solution should be chosen
%-------------------------------------------------------------------------------------
\begin{eqnarray}
\label{randwalk-Zomegaresult} Z = \frac{\sqrt{\omega^2+\kappa_+\kappa_-} - \omega}{2\kappa_-},
\end{eqnarray}
%-------------------------------------------------------------------------------------
since it corresponds to the minimum of ${\cal S}$ over the integration contour.

The expressions presented above provide a link between the spin-chain model in the weak noise limit and
the single-state Markov chain.

\bibliographystyle{spmpsci}
\bibliography{topocurrents}

\end{document}